\documentclass[letterpaper,11pt]{article}
\usepackage{caption}
\usepackage{array}
\pdfoutput=1 

\usepackage[title]{appendix}
\usepackage{extarrows}
\usepackage{amssymb}
\usepackage{amsmath}
\usepackage{setspace}

\usepackage{amstext}
\usepackage{etoolbox}
\let\tinymatrix\smallmatrix

\patchcmd{\tinymatrix}{\scriptstyle}{\scriptscriptstyle}{}{}
\patchcmd{\tinymatrix}{\scriptstyle}{\scriptscriptstyle}{}{}
\patchcmd{\tinymatrix}{\vcenter}{\vtop}{}{}
\patchcmd{\tinymatrix}{\bgroup}{\bgroup\scriptsize}{}{}

\usepackage{graphicx,epsfig}
\usepackage{epsfig}
\usepackage{verbatim} 
\usepackage{color}
\usepackage{ulem}
\usepackage{enumitem}
\usepackage{subfigure}
\usepackage{varwidth,xcolor}
\usepackage{bbm}
\usepackage{parskip}
\usepackage{dsfont}
\usepackage[numbers,sort&compress]{natbib}
\usepackage[body={6.5in, 9in}, right=1in, top=1in]{geometry}
\usepackage{mathtools}
\usepackage{comment}
\usepackage{caption}
\usepackage{bbold}
\usepackage{float}

\newcommand{\Comment}[1]{{}}
\definecolor{darkblue}{rgb}{0.15,0.35,0.55}
\definecolor{darkgreen}{rgb}{0.2,0.7,0.3}
\definecolor{reddish}{rgb}{0.65, 0.2, 0.2}
\usepackage[linktocpage=true]{hyperref}

\newcommand{\be}{\begin{equation}}
\newcommand{\ee}{\end{equation}}
\newcommand{\bea}{\begin{eqnarray}}
\newcommand{\eea}{\end{eqnarray}}
\newcommand{\beas}{\begin{eqnarray*}}
\newcommand{\eeas}{\end{eqnarray*}}

\def\({\left(}
\def\){\right)}

\newcommand{\rd}{{\rm d}}

\def\gsim{ \lower .75ex \hbox{$\sim$} \llap{\raise .27ex \hbox{$>$}} }
\def\lsim{ \lower .75ex \hbox{$\sim$} \llap{\raise .27ex \hbox{$<$}} }

\usepackage{xcolor,colortbl}

\begin{document}
\def\thefootnote{\fnsymbol{footnote}}

\begin{center}
\LARGE{\textbf{Directed Percolation Criticality in Eternal Inflation}} \\[0.5cm]

\large{Justin Khoury and Sam S. C. Wong}
\\[0.5cm]

\small{
\textit{Center for Particle Cosmology, Department of Physics and Astronomy, University of Pennsylvania,\\ Philadelphia, PA 19104}}

\vspace{.2cm}

\end{center}

\vspace{.6cm}

\hrule \vspace{0.2cm}
\centerline{\small{\bf Abstract}}
False-vacuum eternal inflation can be described as a random walk on the network of vacua of the string landscape. In this paper we show that the problem can be mapped naturally to a problem of directed percolation.
The mapping relies on two general and well-justified approximations for transition rates: 1.~the downward approximation, which neglects ``upward" transitions, as these
are generally exponentially suppressed; 2.~the dominant decay channel approximation, which capitalizes on the fact that tunneling rates are exponentially staggered. Lacking detailed knowledge of the string landscape, we model the network of vacua as random graphs with arbitrary degree distribution, including Erd\"os-R\'enyi and scale-free graphs. As a complementary approach, we also model regions of the landscape as regular lattices, specifically Bethe lattices.
We find that the uniform-in-time probabilities proposed in our previous work favor regions of the landscape poised at the directed percolation phase transition. This raises the tantalizing
prospect of deriving universal statistical distributions for physical observables, characterized by critical exponents that are insensitive to the details of the underlying landscape. We illustrate this
with the cosmological constant, and show that the resulting distribution peaks as a power-law for small positive vacuum energy, with a critical exponent uniquely determined by the random graph universality class. 

{\small\noindent 
\vspace{0.3cm}
\noindent
\hrule
\def\thefootnote{\arabic{footnote}}
\setcounter{footnote}{0}

\section{Introduction} 


Our universe appears to be tantalizingly poised at criticality. Extrapolating the Higgs effective potential reveals that the electroweak vacuum lies within a tiny parameter region of metastability~\cite{Degrassi:2012ry,Buttazzo:2013uya,Lalak:2014qua,Andreassen:2014gha,Branchina:2014rva,Bednyakov:2015sca,Iacobellis:2016eof,Andreassen:2017rzq,Khoury:2021zao,Steingasser:2023ugv}, a result that is exquisitely sensitive to the top quark and Higgs boson masses. This hinges on an enormous cancellation between the exponentially small decay rate and the exponentially large observable volume of the universe. It is tempting to speculate that there is an intricate relation between various measured quantities: the cosmological constant (CC), which sets the observable volume of the universe, and the Higgs and top quark masses, which set the Higgs effective potential when extrapolated to high enough energy.

Other fine-tuned features of our universe can also be interpreted as near-criticality. In light of the non-detection of supersymmetry at the electroweak scale, the nearly vanishing ratio of Higgs mass to~$M_{\rm Pl}$ is a result of exponential fine-tuning, which can be interpreted as the boundary of broken/unbroken electroweak symmetry~\cite{Giudice:2006sn}. In cosmology, the CC problem translates to our universe being nearly Minkowski space, which bifurcates into ever-expanding de Sitter (dS)  and crunching Anti-de Sitter (AdS) space-times with distinct asymptotics and stability properties~\cite{Friedrich:1987,Bizon:2011gg}.

In the context of eternal inflation~\cite{Steinhardt:1982kg,Vilenkin:1983xq,Linde:1986fc,Linde:1986fd,Starobinsky:1986fx}, universes in causally-disconnected Hubble patches possess different physical laws. New universes are being constantly generated in all patches. At the same time, string theory predicts an exponentially large number of metastable vacua with enormously rich low-energy physics~\cite{Bousso:2000xa,Kachru:2003aw,Ashok:2003gk}. 
False-vacuum eternal inflation is essentially a random walk on the network of vacua of the string landscape. Given these, it is natural to ask if the near-criticality of our universe can be approached from a statistical point of view. 

In order to extract predictions in the multiverse, it is intuitive to study statistically the distribution of vacua and their associated physical properties. Although deriving such a statistical distribution may ultimately require a complete understanding of quantum gravity, it is still instructive to approximate it using a semi-classical prescription. Attempts to define semi-classical probabilities (or measure) usually rely on limiting frequency distributions. This is perhaps natural, since the infinite ensemble necessary to define frequencies is actually realized in the multiverse. However, it is well known that defining such a measure is ambiguous, as it is assumption-dependent even under the same framework~\cite{Linde:1993nz,Linde:1993xx,Guth:2007ng,Freivogel:2011eg}.

In a recent paper~\cite{Khoury:2022ish}, we presented a general Bayesian framework for probabilistic reasoning in eternal inflation. Different assumptions about the measure problem amount to different choices of priors to define probabilities.
We identified two prior distributions, both pertaining to initial conditions, that must be specified to obtain well-defined occupational probabilities for different vacua. Since eternal inflation is geodesically past-incomplete~\cite{Borde:2001nh}, we know that we exist a finite time~$t$ since the onset of eternal inflation. Our ignorance about the time of existence is captured by a prior density~$\rho(t)$. Relatedly, along our past world-line eternal inflation must have started within some particular ``ancestral" dS vacuum, but we do not know which one. Our ignorance about the ancestral vacuum is parametrized by a probability distribution~$p_\alpha$ over dS vacua. {\it Different proposed solutions to the measure problem simply amount to different choices for these two priors.}

In~\cite{Khoury:2022ish} we argued that there are two natural and well-justified choices for the time-of-existence prior~$\rho(t)$:

\begin{itemize}

\item Since the number of observers grows with volume, a natural choice is~$\rho(t)\sim a^3$. This is equivalent to weighing probabilities by physical volume. The resulting ``late-time/volume-weighted'' probabilities coincide with the measure of Garriga, Vilenkin, Schwartz-Perlov and Winitzki (GSVW)~\cite{Garriga:2005av}. This choice of prior reflects the belief that we exist at asymptotically late times in the unfolding of the multiverse, much later than the exponentially-long relaxation time for the landscape, such that probabilities have settled to a quasi-stationary distribution. This assumption is adopted in nearly all existing approaches to the measure problem~\cite{Linde:1993nz,Linde:1993xx,GarciaBellido:1993wn,Vilenkin:1994ua,Garriga:1997ef,Garriga:2001ri,Garriga:2005av}.

\item Alternatively, motivated by the time-translational invariance of the random walk on the landscape, a natural choice is the uniform prior:~$\rho(t) = {\rm const}$. (To be clear, this is uniform in either proper time or e-folding time.) The resulting ``uniform-in-time" probabilities agree with the prior probabilities of~\cite{Bousso:2006ev}. They are closely related to the ``comoving" probabilities proposed in~\cite{Garriga:2001ri,Garriga:2005av}, as well as probabilities derived recently using the local Wheeler-De Witt equation~\cite{Friedrich:2022tqk}. Importantly, the uniform-in-time probabilities favor vacua that are accessed early on in the evolution of the multiverse, during the {\it approach to equilibrium}. This is consistent with the early-time approach to eternal inflation developed recently~\cite{Denef:2017cxt,Khoury:2019yoo,Khoury:2019ajl,Kartvelishvili:2020thd,Khoury:2021grg}. 

\end{itemize}

The Bayesian framework allowed us to compare the plausibility of the uniform-in-time and late-time measures to explain our data
by computing the Bayesian evidence for each. We argued, under general and plausible assumptions, that posterior odds overwhelmingly favor uniform-in-time probabilities~\cite{Khoury:2022ish}.
The argument, briefly reviewed in Sec.~\ref{RW complex nets}, relies on assumptions that have been made in previous studies of the landscape. There are some caveats, of course, and we tried
to enunciate them carefully in~\cite{Khoury:2022ish}. For this reason, we believe that the uniform-in-time measure is the correct objective approach to probabilistic reasoning in the multiverse.
In this work we therefore focus on uniform-in-time probabilities.

\subsection{Directed percolation in eternal inflation}

Our goal in this paper is to show how vacuum dynamics on the landscape can be mapped naturally to a problem of directed percolation. The mapping relies on two very general and reasonable assumptions about
transition rates between vacua:

\begin{enumerate} 

\item The first assumption is that transition rates between dS vacua satisfy a condition of detailed balance, such that ``upward" jumps are exponentially suppressed by~$\frac{\kappa_{\rm up}}{\kappa_{\rm down}} \sim {\rm e}^{-\Delta S}$,
where~$S$ is the dS entropy. This is satisfied by most tunneling instantons, including Coleman-De Luccia (CDL)~\cite{Coleman:1977py,Callan:1977pt,Coleman:1980aw}. Thus we are justified to work in the {\it downward  approximation}~\cite{SchwartzPerlov:2006hi,Olum:2007yk}, wherein upward transitions are neglected to leading order. In this approximation, the network of vacua becomes a {\it directed graph}. 

\item The second assumption rests on the fact that semi-classical tunneling rates depend exponentially on the Euclidean action of the instanton,~$\kappa\sim {\rm e}^{-S_{\rm E}}$. In turn,~$S_{\rm E}$ depends sensitively on the height and width of the potential barrier. Because of this exponential sensitivity, branching ratios for dS vacua are typically overwhelmingly dominated by a single decay channel. This defines the {\it dominant decay channel}, in which exponentially-subdominant decay channels are neglected.  

\end{enumerate}

It should be stressed that these approximations are not strictly necessary to study percolation. They are made for convenience, to simplify the problem, and we will discuss how the analysis can be generalized by relaxing them.
In any case, with these approximations, the uniform-in-time probabilities reduce to a simple and intuitive observable in directed graphs. Namely, the probability to occupy a given node~$I$ simplifies to
\be
P(I) \sim s_I\,,
\label{PIsI intro}
\ee
where~$s_I$ is the number of ancestors, {\it i.e.}, nodes that can reach~$I$ through a sequence of directed (downward) transitions. See Fig.~\ref{number anc fig}. Thus the measure favor
vacua with a large basin of ancestors. In other words, {\it regions of the landscape with large probability must therefore have the topography of a deep valley, or funnel}~\cite{Khoury:2019yoo,Khoury:2019ajl,Khoury:2021grg,Khoury:2022ish}.
This is akin to the smooth folding funnels of protein conformation landscapes~\cite{proteins1}, and those of atomic clusters with Leonard-Jones interactions~\cite{Doye2002,Doye2004,Doye2005}. In the context of deep learning, it has been argued that deep neural networks that generalize well have a loss function characterized by a smooth funnel~\cite{DDNlossfunnel}. Another instance is the ``big valley'' hypothesis in combinatorial optimization ({\it e.g.}, the search space of the traveling salesman problem), where it is conjectured that local optima are clustered around the central global optimum~\cite{TSP}. It is tempting to speculate that funnels are a generic solution to optimization problems on complex energy landscapes.

Equation~\eqref{PIsI intro} gives an intuitive and well-justified notion of probability for different vacua. But what can we reasonably assume about the network of vacua, given our limited understanding of the string landscape? Lacking detailed knowledge of the underlying network of vacua, it seems sensible to model regions of the landscape as random graphs. Random graphs have a long and venerable history, going back to the seminal work of Erd\"os and R\'enyi graph~\cite{Erdos:1959:pmd}. In general, they can be defined by specifying a probability distribution that a given node has a certain degree (number of links), with Erd\"os-R\'enyi graphs corresponding to the special case of a Poisson degree distribution. In this work we follow~\cite{PhysRevE.64.026118} and consider arbitrary degree distributions, including scale-free random graphs. 

As a complementary approach, we also model landscape regions as a regular lattice, specifically a Bethe lattice (or Cayley tree). This is suitable for local string landscapes in which vacua form a regular network, for instance the axion landscape~\cite{Bachlechner:2015gwa,Bachlechner:2017zpb,Bachlechner:2018gew}. The directed percolation transition can be studied analytically for both Bethe lattices and (Erd\"os-R\'enyi) random graphs~\cite{PhysRevE.64.026118}. 
Remarkably, despite being extreme opposites in terms of graph ``regularity", Bethe lattices and Erd\"os-R\'enyi graphs belong to the same percolation universality class. Thus it is our hope that, despite being highly simplified and idealized, these two approaches offer important lessons about percolation phenomena on the landscape, that are applicable to more realistic dynamics.

\begin{figure}
\begin{center}
    \includegraphics[width=0.9\textwidth]{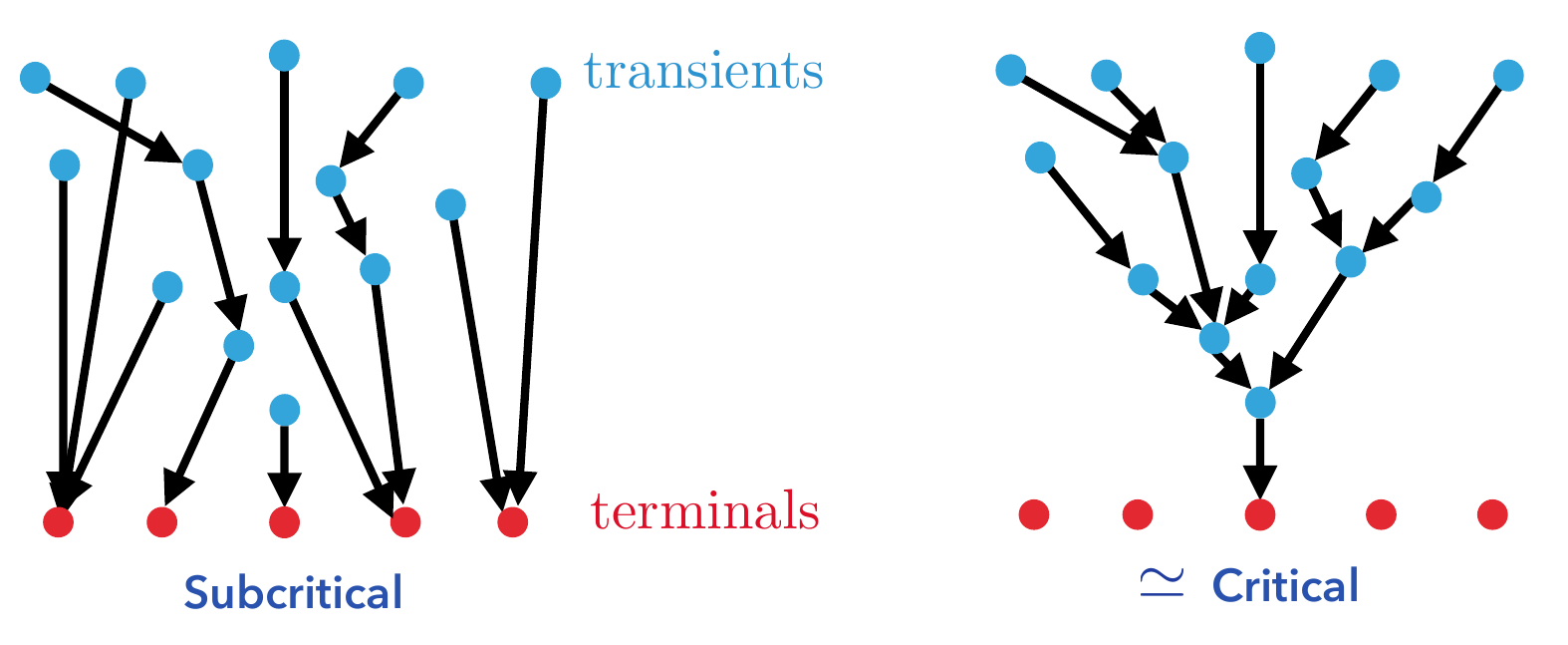}
  \caption{\label{percolation sketch} A region of the landscape, comprised of transient dS vacua (blue nodes) and terminals (red nodes). In the dominant decay channel approximation, each dS vacuum has exactly one outgoing link. {\it Left:} If dS vacua decay primarily to terminals, the region breaks up into many small disconnected components and is therefore subcritical. {\it Right:} If dS vacua mainly decay to other dS vacua, then a giant connected component can emerge, and the region is near percolation criticality.}
\vspace{-5pt}  
\end{center}
\end{figure}

To see how this maps to directed percolation, consider the landscape networks shown in Fig.~\ref{percolation sketch}, comprised of dS transient vacua (blue nodes) and AdS/Minkowski terminal vacua (red nodes).\footnote{We assume throughout that AdS/Minkowski vacua are terminal, acting as absorbing nodes. See~\cite{Cespedes:2023jdk}, however, for a recent discussion of the possibility to up-tunnel out of AdS.} In the downward approximation, all transitions are one-way, and the graphs are directed. Each dS transient has exactly one directed edge emanating from it, corresponding to its dominant decay channel. Now, if dS vacua mainly decay to terminals (left panel), then the region breaks up into many small disconnected components, resulting in~$s_I \sim {\cal O}(1)$, and thus low probability. If, however, dS vacua mainly decay to other dS vacua (right panel), then the graph can include a very large connected component, with low-lying nodes having~$s_I \gg 1$. This corresponds to the emergence of a giant component at the percolation phase transition. 

It is clear from these simple considerations that {\it vacua with the highest occupational probability reside in landscape regions poised at the directed percolation phase transition.}
This is a key result of our analysis. In hindsight, since the uniform-in-time measure is relevant for the approach to equilibrium in landscape dynamics, its relation to directed percolation
is perhaps not surprising, as directed percolation is the paradigmatic non-equilibrium critical phenomenon~\cite{Odor:2002hk}. 

As usual, the power of criticality lies in universality. Near the percolation phase transition, various quantities assume power-law (scale-invariant) probability distributions, characterized by
universal critical exponents that are insensitive to the microscopic details of the system. This raises the tantalizing possibility of deriving universal probability distributions for physical
observables without detailed knowledge of the underlying landscape. We illustrate this point concretely with the CC, and briefly mention other potential observables in the conclusions.

\subsection{Universal probability distribution for the CC}

As reviewed in Sec.~\ref{sec:percolation}, at criticality the probability~$P_s$ that a randomly-chosen vacuum has~$s$ ancestors displays a power-law tail,~$P_s \sim s^{-3/2}$. (The particular critical exponent of~$-3/2$ holds for the Erd\"os-R\'enyi universality class; scale-free graphs have different critical exponents.) Assuming only that the underlying CC probability distribution on the landscape is smooth as~$v\rightarrow 0^+$, we show in Sec.~\ref{CC distribution} that this translates into a (non-anthropic) universal probability distribution for the CC (which takes into account eternal inflationary dynamics) that is also power-law near the origin:
\be
P(v) \sim v^{-3/2} \,. 
\label{CC dis intro}
\ee
Thus the uniform-in-time cosmological measure favors small, positive vacuum energy. Quantitatively, the~$95\%$ confidence interval for the CC is set by the size of the giant component, which
is famously~${\cal O}\big(N_{\rm dS}^{2/3}\big)$ for Erd\"os-R\'enyi graphs:
\be
v\lesssim N^{-2/3}_{\rm dS} \,.
\ee
This can explain the observed CC,~$v_{\rm obs} \sim 10^{-120}$, if our vacuum belongs to a funneled region of size~$N_{\rm dS}\sim 10^{240}$. To be clear, here~$N_{\rm dS}$ is the number of dS (transient) vacua in a funnel region near directed percolation criticality, not the total number of dS vacua across the entire landscape. Since the measure favors vacua with the largest number of ancestors, we are likely to inhabit the largest funnel region near percolation criticality, {\it i.e.}, the near-critical region with largest~$N_{\rm dS}$. 

Before closing, we should mention other occurrences of percolation criticality in the context of eternal inflation, and how they contrast with our framework. It is well known that the bubbles generated in false-vacuum inflation exhibit a percolation phase transition when the nucleation probability within a Hubble volume,~$\kappa\equiv \Gamma/H^4$, reaches a critical value somewhere in the range~$10^{-6} \lesssim \kappa_{\rm c} \lesssim 0.24$~\cite{Guth:1982pn}. This transition describes bubble percolation in space-time, as opposed to the percolation phase transition in the network of vacua discussed in this work. Relatedly, in the context of slow-roll inflation, it was shown in~\cite{Creminelli:2008es} that the phase transition to eternal inflation can be described by a Galton-Watson branching process~\cite{branching}, whose critical behavior is equivalent to directed percolation, {\it e.g.},~\cite{bollobasriordan}. Lastly, we should mention the mechanism of `self-organized localization'~\cite{Giudice:2021viw}, whereby the near-criticality of our universe arises from quantum first-order phase transitions in stochastic inflation. In contrast, our approach pertains to classical, second-order non-equilibrium criticality.

The paper is organized as follows. In Sec.~\ref{RW complex nets} we briefly review vacuum dynamics as an absorbing Markov process on the network of vacua. In Sec.~\ref{Bayes sec} we describe the general Bayesian approach to the measure problem, and review the late-time/volume-weighted and uniform-in-time probabilities as two well-justified choices of priors. We also review the argument, originally given in~\cite{Khoury:2022ish}, that posterior odds overwhelmingly favor the uniform-in-time hypothesis. In Sec.~\ref{DP map}, we discuss how the mapping of vacuum dynamics to a problem of directed percolation. Section~\ref{sec:percolation} is a rather comprehensive review of the key notions of percolation on random graphs, both undirected and directed, with general degree distributions. In Sec.~\ref{DP eternal infn} we apply these notions to the case of interest, namely random networks with terminal (AdS) vacua, and argue that uniform-in-time probabilities favor regions of the landscape poised at directed percolation criticality. In Sec.~\ref{CC distribution} we show how the probability distributions of ancestors and descendants, which assume power-law tails at criticality, translate into universal distributions for the CC with certain critical exponents. We summarize our results and discuss a few avenues of future research in Sec.~\ref{sec:concl}.

\section{Brief Review of Vacuum Dynamics}
\label{RW complex nets}

Vacuum dynamics on the string landscape are described by a linear Markov process~\cite{Garriga:1997ef,Garriga:2005av}. Technically, this is an absorbing Markov process, because of AdS vacua which act as terminals. 
As a result, detailed balance is explicitly violated, and the dynamics are out of equilibrium. The Markov process describes the probability~$f_I(t)$ along a given world-line to occupy vacuum~$I$ as a function of time.
This probability satisfies the master equation 
\be
\frac{{\rm d}f_I}{{\rm d}t} = \sum_{J} \kappa_{IJ}f_{J}  - \sum_K \kappa_{KI} f_{I}  \,,
\label{master}
\ee
where~$\kappa_{IJ}$ is the $J \rightarrow I$ transition rate. (Terminal vacua by definition have~$\kappa_{a I} = 0$.) While most of the results in this section hold for general tunneling rates, we have in mind transitions mediated by semi-classical instantons, such as Coleman-De Luccia (CDL)~\cite{Coleman:1977py,Callan:1977pt,Coleman:1980aw}, Hawking-Moss~\cite{Hawking:1981fz} and Brown-Teittleboim~\cite{Brown:1987dd}. 

The general time variable~$t$ is related to proper time in vacuum~$I$ via a lapse function:
\be
\Delta \tau_I = {\cal N}_I \Delta t\,. 
\ee
The master equation relies on coarse-graining over a time~$\Delta \tau_I$, which must be longer than any transient evolution between epochs of vacuum energy domination.\footnote{At the same time,~$\Delta \tau_I$ should be shorter than the
 lifetime of most metastable dS vacua, for otherwise we would be ``integrating out" the transitions we are interested in describing. In practice, the coarse-graining time interval for a given transition to~$I$ should satisfy~$\Delta\tau_I \gtrsim |H_I|^{-1} \log \frac{H_{\rm parent}}{|H_I|}$, where~$H_{\rm parent}$ is the Hubble rate of the parent dS vacuum (see, {\it e.g.},~\cite{Salem:2012wa}).} Since AdS bubbles crunch in a Hubble time, coarse-graining spans their entire evolution.

The probabilities~$f_I(t)$ have a dual interpretation. They can be interpreted ``locally'', as occupational probabilities along a world-line. Or, they can be interpreted ``globally", as the fraction of comoving volume that each vacuum occupies on a spatial hypersurface at time~$t$. We mainly adopt the former interpretation. Equation~\eqref{master} makes two properties of the~$f_I$'s clear:~$1)$~The master equation~\eqref{master}
is manifestly invariant under redefinitions of~$t$, hence the~$f_I$'s are time-reparameterization invariant; $2)$~Because summing the right-hand side over~$I$ gives zero, the~$f_I$'s
can be normalized:
\be
\sum_I f_I = 1\,.
\ee 
Thus the~$f_I(t)$'s offer well-defined, time-reparametrization invariant probabilities to occupy different vacua at time~$t$.

We will be primarily interested in the dS component of the master equation, given by
\be
\frac{{\rm d}f_i}{{\rm d}t}  = \sum_j M_{ij} f_j  \,;\qquad M_{ij}\equiv \kappa_{ij} - \delta_{ij} \kappa_j\,.
\label{master dS}
\ee
Here,~$M_{ij}$ is the~${\rm dS}\rightarrow {\rm dS}$ transition matrix, and~$\kappa_i \equiv \sum_J \kappa_{Ji}$ is the total decay rate of vacuum~$i$. Our only assumption about~$M_{ij}$ is that it is irreducible, {\it i.e.}, there exists a sequence of transitions connecting any pair of dS vacua, a property which has been argued to be valid for the string landscape~\cite{Brown:2011ry}. Equation~\eqref{master dS} can be solved in terms of a Green's function:
\be
f_i(t) = \sum_\alpha \left({\rm e}^{Mt}\right)_{i\alpha} p_\alpha\,, 
\label{f soln}
\ee
where~$p_\alpha \equiv f_\alpha(0)$ is the initial probability over ancestral vacua. Since eternal inflation by definition started in a dS vacuum, the initial probabilities satisfy
\be
\sum\limits_{\alpha = 1}^{N_{\rm dS}} p_\alpha = 1\,.
\ee

\section{Bayesian Probabilities}
\label{Bayes sec}

To define Bayesian probabilities, one must carefully distinguish the elements that are inherent to the eternal inflation hypothesis from those 
that require additional assumptions in the form of prior information. An important fact is that eternal inflation, while eternal into the future, is not 
eternal into the past. That is, an eternally-inflating space-time is past geodesically incomplete~\cite{Borde:2001nh}. This has two implications: 

\begin{itemize}

\item We exist a finite time~$t$ after the onset of eternal inflation, but we do not know how long ago that was. We must therefore parametrize our ignorance about the time of existence with a prior density~$\rho(t)$, normalized as~$\int_0^\infty {\rm d}t \,\rho(t) = 1$.

\item Along our past world-line, eternal inflation started in some ancestral dS vacuum~$\alpha$, but we do not know which one. Our ignorance about the ancestral vacuum is captured by the initial probability distribution~$p_\alpha$.

\end{itemize}

\noindent Lastly, it is customary to condition probabilities on one piece of observational data. Namely, that we exist in our dS pocket universe during the transient period before vacuum domination. That is,
we exist within a coarse-graining time~$\Delta t$ after nucleation of our bubble.

It is then straightforward to write down the joint probability distribution~$P(I,t,\alpha)$ to inhabit a bubble of vacuum~$I$, nucleated at time~$t$, starting from an ancestral vacuum~$\alpha$:
\be
P(I,t,\alpha) \sim \sum_{j} \kappa_{Ij}\Delta t \,  \left({\rm e}^{Mt}\right)_{j\alpha} p_\alpha \rho(t) \,.
\ee
This is easy to understand. The factor~$\left({\rm e}^{Mt}\right)_{j\alpha} p_\alpha$ is the probability to evolve from ancestral vacuum~$\alpha$ to parent dS vacuum~$j$ at time~$t$; while the factor~$\kappa_{Ij}\Delta t$ is the probability to transition from parent dS vacuum~$j$ to vacuum~$I$ in the next~$\Delta t$. Lastly, we weigh the time of nucleation with~$\rho(t)$, and sum over all dS parents~$j$. To our mind, the above joint probabilities are the correct objective
approach to inductive reasoning in the multiverse. They accurately encode our ignorance about when and where eternal inflation started in our past. {\it Different approaches to the measure problem simply amount to different choices for the priors~$p_\alpha$ and~$\rho(t)$.} 

Within this general framework, one can perform the three main operations of Bayesian inference:

\begin{enumerate}

\item By marginalizing over the model parameters~$t$ and~$\alpha$, and using~\eqref{f soln}, we obtain the {\it prior predictive distribution}:
\be
P(I) \sim \sum_{j} \kappa_{Ij}\Delta t \int_0^\infty {\rm d} t \, f_j(t) \rho(t) \,.
\label{PI}
\ee
This distribution informs us on which vacua are statistically favored without taking any data ({\it e.g.}, value of the CC, particle spectrum {\it etc.}) into consideration, other than conditioning on our bubble being nucleated within the last~$\Delta t$. 

\item Different hypotheses~${\cal H}_1$ and~${\cal H}_2$, corresponding to different choices of priors, can be compared by computing the posterior odds:
\be
\frac{P({\cal H}_1 | D)}{P({\cal H}_2 | D)} = \frac{P(D | {\cal H}_1)}{P(D | {\cal H}_2)}\frac{P({\cal H}_1)}{P({\cal H}_2)}\,,
\label{Bayes factor}
\ee
where~$P({\cal H}_i)$ is the prior odds for each hypothesis, and~$\frac{P(D | {\cal H}_1)}{P(D | {\cal H}_2)}$ is the Bayes factor. The data~$D$ refers to all the information available about our observable universe, in the form of measured values for various observables~$\{O_i\}$. These include the particle content, masses and couplings of the Standard Model, as well as the parameters of the cosmological~$\Lambda$CDM model.

\item Conditioning on our data~$D$ for a given choice of priors, we can perform parameter inference. For instance,~$P(t|D)$ gives the posterior distribution for the time of nucleation.

\end{enumerate}

\noindent Each of these operations was studied in detail in~\cite{Khoury:2022ish}. In what follows we will be primarily interested in the prior predictive probabilities~\eqref{PI}.

\subsection{Uniform-in-time measure}

As mentioned above, a choice of measure amounts to specifying a choice of priors~$p_\alpha$ and~$\rho(t)$. Consistency requires that priors reflect {\it all} information at hand, while at the same time being minimally informative. 

Let us first discuss the time of nucleation prior~$\rho(t)$, as it is most important to determine~$P(I)$. In general, specifying a prior for a continuous variable is tricky, for the obvious reason that a uniform prior is not reparametrization invariant. Following Jaynes~\cite{jaynes03}, a useful strategy in this case is to identify the symmetries of the problem and apply the notion of group invariance. Logical consistency requires that our prior be invariant under all symmetry transformations.

In the case at hand, a key property of the master equation~\eqref{master} is that it is {\it time-translation invariant}. More precisely it is invariant under translations in proper time, as well as 
any time variable~$t$ related to proper time via a lapse function~${\cal N}_I$ depending on~$H_I$ only ({\it e.g.}, scale factor/e-folding time). Without additional information, the most natural choice is the uniform prior:\footnote{Of course, a uniform distribution on the half real line is not normalizable, so~\eqref{uni} is technically an improper prior. One can instead work with a regularized prior, such as~$\rho(t) = \epsilon {\rm e}^{-\epsilon t}$ or~$\rho(t) = 1/T$ over~$0 \leq t \leq T$. As shown in~\cite{Khoury:2022ish}, the resulting probabilities are independent of the regulator as it is removed, {\it i.e.}, as~$\epsilon \rightarrow 0$ or~$T\rightarrow\infty$, respectively.}
\be
\rho(t) = {\rm constant}\,.  
\label{uni}
\ee
To be clear, this prior is uniform in proper time and e-folding time. In terms of conformal time, however, it corresponds to the Jeffreys prior,~$\rho(\eta) \sim \eta^{-1}$, consistent with the dS dilation symmetry~$\eta \rightarrow \lambda \eta$,~$\vec{x}\rightarrow\lambda\vec{x}$. 

Substituting into~\eqref{PI}, we can perform the time integral using the identify
\be
\int_0^\infty {\rm d}t \left({\rm e}^{Mt}\right)_{ij} = -M^{-1}_{ij}  = \kappa_i^{-1} \left(\mathds{1} - T\right)^{-1}_{ij} \,,
\label{Laplace}
\ee
where~$T_{ij} \equiv  \frac{\kappa_{ij}}{\kappa_j}$ is the branching ratio. (More generally, the branching ratio matrix has components~$T_{Ij} = \frac{\kappa_{Ij}}{\kappa_j}$,~$T_{ab} = \delta_{ab}$, and~$T_{ja} = 0$, such that~$\sum\limits_{I} T_{IJ}=1$ for all~$J$.) Equation~\eqref{PI} then gives
\be
P(I) \sim \sum_j T_{Ij} \sum_\alpha  \big(\mathds{1} - T \big)^{-1}_{j\alpha} p_\alpha  \,.
\label{Puni p_alpha}
\ee
The matrix~$\left(\mathds{1} - T\right)^{-1}$ is known as the fundamental matrix for the absorbing Markov chain. Expanding it as a geometric series,~$\left(\mathds{1} - T\right)^{-1}_{ij} =\delta_{ij} + T_{ij} + \sum_k T_{ik}T_{kj} + \ldots$, it is easily recognized as the total branching probability for all transition paths connecting~$j$ to~$i$. Thus~$P(I)$ naturally interpreted as the sum over all paths connecting ancestral vacua to vacuum~$I$, weighted by the branching probability for each path and averaged over ancestral vacua.

Next, consider the prior~$p_\alpha$ over ancestral vacua. This was discussed in detail in~\cite{Khoury:2022ish}, and we briefly mention the salient points. The prior~$p_\alpha$ pertains to the question of the initial state in quantum cosmology, which has been the subject of active debate for decades and remains an open problem. A well-motivated proposal for the quantum creation of a closed universe is the Hartle-Hawking (HH) state~\cite{Vilenkin:1982de,Hartle:1983ai}, which exponentially favors the lowest energy (highest entropy) dS vacuum. Another well-studied proposal is the tunneling wavefunction~\cite{Linde:1983mx,Linde:1983cm,Vilenkin:1984wp,Vilenkin:1986cy}, which instead favors high-energy/low-entropy
initial vacua. Thus the tunneling wavefunction favors (high-energy) inflation, whereas the HH state does not~\cite{Vilenkin:1987kf}. 

As motivated in~\cite{Khoury:2022ish}, a reasonable attitude is to err on the side of maximal ignorance and apply the principle of indifference:
\be
p_\alpha = \frac{1}{N_{\rm dS}}\,.
\label{anc uni}
\ee
(If the number of dS vacua in the landscape is infinite~\cite{Silverstein:2001xn,Maloney:2002rr,DeLuca:2021pej}, then~\eqref{anc uni} would represent an improper prior, which is fine since the resulting probabilities would nevertheless be well-defined.)
Because high-energy dS vacua are expected to vastly outnumber low-energy dS vacua in the landscape, a uniform prior is statistically equivalent to a prior favoring high-energy/low entropy
initial conditions, such as the tunneling wavefunction. If the HH state turns out to be the correct initial conditions for eternal inflation in our past, then this
would have important implications for the uniform-in-time probabilities.

Adopting~\eqref{anc uni},~\eqref{Puni p_alpha} reduces to
\be
\boxed{P(I) \sim \sum_j T_{Ij} \sum_\alpha  \big(\mathds{1} - T \big)^{-1}_{j\alpha}\,.}
\label{Puni}
\ee
This distribution agrees with the prior probabilities of~\cite{Bousso:2006ev}, and is closely related to the ``comoving" probabilities proposed in~\cite{Garriga:2001ri,Garriga:2005av}.
We will see that this admits a clear and intuitive interpretation with the simplifying assumptions discussed in Sec.~\ref{DP map}.

\subsection{Late-time/volume-weighted measure}

Another reasonable choice for~$\rho(t)$ is motivated by the fact that the number of observers grows with volume. Hence,~$\rho(t)$ should grow accordingly:\footnote{For the prior distribution to be normalizable, a regulator is once again necessary. This can be achieved simply by imposing a cutoff time~$t_{\rm c}$, and letting~$t_{\rm c}\rightarrow\infty$ at the end of the calculation. The resulting probabilities are insensitive to the cutoff.}
\be
\rho(t) \sim a^3(t)\,.
\ee
As shown in~\cite{Khoury:2022ish}, this is equivalent to weighing occupational probabilities by physical volume. 

Because the prior is sharply peaked at late times, the occupational probabilities~$f_j(t)$ can be approximated by their asymptotic form
\be
f_j(t) \simeq s_j {\rm e}^{-q t}  \,.
\label{fj late}
\ee
Here~$s_j$ the so-called dominant eigenvector of~$M_{ij}$, which by definition has the largest (least negative) eigenvalue~$-q$~\cite{Garriga:2005av}.
Substituting into~\eqref{PI}, we obtain in this case
\be
P_{\rm late}(I) \sim \sum\limits_j \kappa_{Ij} s_j \,.
\label{PIlate}
\ee
This agrees with the GSVW measure~\cite{Garriga:2005av} obtained by counting bubbles along a world-line.

The above distribution admits an intuitive explanation in downward perturbation theory, discussed in Sec.~\ref{down sec} below. In this approximation, the dominant eigenvector takes a simple form~\cite{Olum:2007yk}:
\be
s_j \simeq \frac{\kappa_\star}{\kappa_j} \left(\mathds{1} - T\right)^{-1}_{j\star} \,,
\label{s simple}
\ee
where~$\star$ denotes the most stable ({\it i.e.}, longest-lived) dS vacuum, also known as the dominant vacuum. Thus~\eqref{PIlate} becomes
\be
P_{\rm late}(I) \sim \sum\limits_j T_{Ij} \left(\mathds{1} - T\right)^{-1}_{j\star} \,.
\label{Plate}
\ee
In other words, this is recognized as the total branching probability from~$\star$ to~$I$. The late-time/volume-weighted measure is independent of initial conditions ({\it i.e.}, independent of~$p_\alpha$),
reflecting the attractor nature of eternal inflation. However, somewhat paradoxically,~\eqref{Plate} coincides with~\eqref{Puni  p_alpha} for the special choice of initial conditions~$p_\alpha = \delta_{\alpha \star}$.

\subsection{Model comparison favors uniform-in-time probabilities}
\label{model comp}

In~\cite{Khoury:2022ish} we compared the Bayesian evidence for the uniform-in-time and late-time measures by computing the Bayes factor~$\frac{P(D | {\cal H}_{\rm late})}{P(D | {\cal H}_{\rm uni})}$.  
We argued, under general and plausible assumptions, that it overwhelmingly favors the uniform-in-time hypothesis. The reason is easily understood intuitively. Since~$\star$ is the most stable
vacuum anywhere in the landscape, it is likely that it can only decay via an upward transition, because upward jumps are doubly-exponentially suppressed (as discussed in Sec.~\ref{down sec}). Therefore,
the branching probability to vacua compatible with our data is also doubly-exponentially suppressed. In contrast, for uniform-in-time probabilities, if vacua compatible with our data can be reached from
some ancestral vacua via a sequence of downward transitions, then the Bayesian evidence is likely exponentially small, but not doubly-exponentially suppressed. 

Furthermore, conditioning on our data~$D$, we performed in~\cite{Khoury:2022ish} parameter inference to determine the most likely time of nucleation. For the uniform-in-time hypothesis, we found that the average time for occupying vacua compatible with our data is much shorter than mixing time for the landscape. This is fully consistent with the ``early-time" approach to eternal inflation~\cite{Khoury:2019yoo,Khoury:2019ajl,Kartvelishvili:2020thd,Khoury:2021grg},
which proposes that we live during the {\it approach to equilibrium} in the unfolding of the multiverse. See~\cite{Denef:2017cxt} for related ideas. This is in contrast with the late-time/volume-weighted distribution, 
which reflects the belief that the evolution of the multiverse has been going on for an exponentially long time, much longer than the mixing time of the landscape, such that the occupational probabilities have settled to a quasi-stationary distribution.

We henceforth focus on the uniform-in-time measure~\eqref{Puni}. 

\section{Mapping to a Directed Percolation Problem}
\label{DP map}

The landscape can be modeled as a network (or graph). The nodes represent the different vacua, while the links define the network topology and represent all relevant transitions between vacua.
There are two types of nodes: transients (dS) and terminals/absorbing (AdS and Minkowski). Because transition rates are different along each link, the graph is said to be weighted. The master equation~\eqref{master} 
describes a random walk on this weighted network. The measures derived above are closely related to network centrality indices: the uniform-in-time measure~\eqref{Puni} is analogous to Katz centrality~\cite{Katz1953}; the late-time (GSVW) measure~\eqref{Plate} to eigenvector centrality. 

In this Section we show how the problem can be mapped to a problem of directed percolation. Directed percolation is the paradigmatic critical phenomenon for non-equilibrium systems~\cite{Odor:2002hk}. It is perhaps not surprising that the
absorbing Markov process describing vacuum dynamics on the landscape, which is inherently non-equilibrium, belongs to the universality class of directed percolation. The mapping relies on two very general and reasonable assumptions about transition rates between vacua, discussed respectively in Secs.~\ref{down sec} and~\ref{dom sec}. We will see that, with these approximations, the uniform-in-time probabilities reduce to a simple and intuitive observable in directed graphs, namely the number of ancestors of a given node.

\subsection{Downward approximation}
\label{down sec}

The first assumption is that transitions between dS vacua satisfy a condition of {\it detailed balance}~\cite{Lee:1987qc}:
\be
\frac{\kappa_{ji}}{\kappa_{ij}}  \sim {\rm e}^{S_j-S_i}\,,
\label{detailed balance}
\ee
where~$S_j =  8\pi^2M_{\rm Pl}^2/H_j^2$ is the dS entropy. This condition is satisfied by CDL, Hawking-Moss and Brown-Teittleboim instantons. More generally, it is consistent with the interpretation
of quantum dS space as a thermal state~\cite{Dyson:2002pf}.\footnote{Notably,~\eqref{detailed balance} is violated by the Farvi-Guth-Guven process~\cite{Farhi:1989yr}, though the interpretation of its singular instanton is still unsettled~\cite{Fischler:1989se,Fischler:1990pk,DeAlwis:2019rxg,Fu:2019oyc,Mirbabayi:2020grb}. It is also violated by the mechanism of nucleating localized, high-energy regions proposed recently~\cite{Olum:2021pux}. Upward jumps are still suppressed in this case, by at least~${\rm e}^{\# M_{\rm Pl}/H_{\rm low}}$ instead of~${\rm e}^{\# M_{\rm Pl}^2/H_{\rm low}^2}$.}
Notice that~\eqref{detailed balance} depends only on the false and true vacuum potential energy --- it is insensitive to the potential barrier and does not rely on the thin-wall approximation. 

Equation~\eqref{detailed balance} implies that upward transitions, which increase the potential energy, are exponentially suppressed compared to downward tunneling.
This allows one to define a ``downward" approximation~\cite{SchwartzPerlov:2006hi,Olum:2007yk}, in which upward transitions are neglected to zeroth order. (Upward transitions are treated perturbatively at higher order.) In this approximation, the
network of vacua reduces to a directed, acyclic graph~\cite{DAGpaper}, {\it i.e.}, without directed loops, whereby a link from~$j$ to~$i$ is only allowed if~$V_j \geq V_i$. This may be a good place to point out that 
the validity of the master equation~\eqref{master} has not been rigorously established for upward transitions. So it may be the case that the description of vacuum dynamics as a Markov process is only legitimate in the strict downward approximation.

In any case, dS vacua that can only decay via upward transitions become effectively terminal in this approximation. In other words, in the downward approximation terminals
consist both of AdS/Minkowski vacua and dS vacua with upward-only decay channels. Transient nodes are dS vacua with at least one downward decay channel.

\begin{figure}
\begin{center}
    \includegraphics[width=0.75\textwidth]{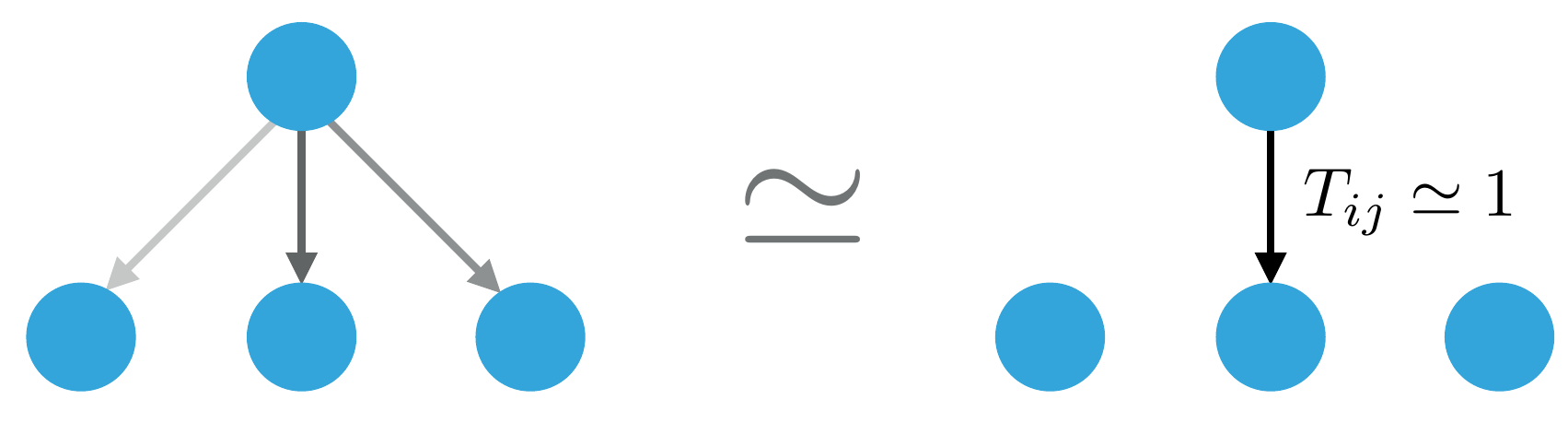}
  \caption{\label{branch ratio fig} This explains the dominant decay channel approximation. On the left, in the downward approximation a given node has possibly many allowed decay channels (directed links), but with exponentially staggered branching ratios (different shades of gray). On the right, the approximation amounts to only keeping the link with largest branching ratio. The parent node therefore has out-degree~1.}
\end{center}
\end{figure}

\subsection{Dominant decay channel}
\label{dom sec}

The second assumption is motivated by a generic feature of transition rates in quantum field theory, namely that they are exponentially staggered. This is because tunneling rates depend exponentially on the instanton Euclidean action:
\be
\kappa\sim {\rm e}^{-S_{\rm E}}\,.
\ee
For CDL tunneling, for instance,~$S_{\rm E}$ depends sensitively on the shape of the potential, such as the height and width of the barrier. Because of this exponential sensitivity,
branching ratios for dS vacua are typically overwhelmingly dominated by a single decay channel, with~$T_{Ij}^{\rm dom} \simeq 1$, while other decay channels are comparatively exponentially suppressed,
 {\it i.e.},~$T_{Ij}^{\rm other} \simeq 0$. 

Hence our second simplifying assumption is that we work in the approximation where~$T_{Ij}$ is 0 or 1. In other words, either there is a link between two nodes~($T_{Ij} \simeq 1$) or not~($T_{Ij} \simeq 0$). Furthermore, if~$T_{Ij} \simeq 1$, then this is the only link emanating from~$j$, {\it i.e.}, the out-degree of~$j$ is~1. This is illustrated in Fig.~\ref{branch ratio fig}.  There are of course exceptions, for instance in regular lattices of flux vacua~\cite{Bousso:2000xa}. But we expect that single-channel dominance is justified for random landscapes, which will be our primary interest in Sec.~\ref{sec:percolation}.

\begin{figure}
\begin{center}
    \includegraphics[width=1.0\textwidth]{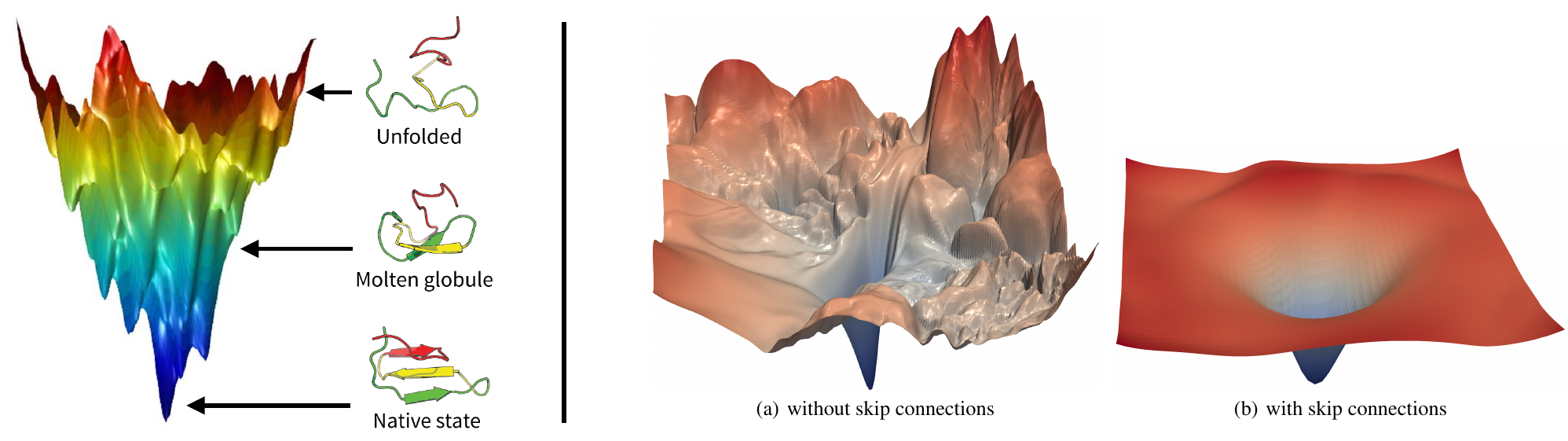}
  \caption{\label{funnel} The uniform-in-time probabilities favor regions of the landscape with the topography of a deep valley, or funnel. This is akin to the free energy landscapes
  of naturally-occurring proteins ({\it Left}, reproduced from~\cite{funnelfig}), which are characterized by a large funnel near the native state. {\it Right:} A similar narrative holds in deep learning. This shows the loss surfaces of ResNet-56 with/without skip connections, reproduced from~\cite{DDNlossfunnel}. Skip connections lead to better generalization, and correspond to a loss function characterized by a smooth funnel.}
\vspace{-10pt}  
\end{center}
\end{figure}

\subsection{Implications for uniform-in-time probabilities}

The two assumptions discussed above greatly simplify the uniform-in-time probabilities~\eqref{Puni}, and lead to an intuitive interpretation.

\begin{itemize}

\item {\bf Downward approximation:} In this approximation the only contributing paths to a given vacuum~$I$ are those composed of a sequence of downward transitions.
It follows that the probabilities~\eqref{Puni} favor vacua that can be accessed through downward transitions from a large basin of ancestors.
In other words, {\it regions of the landscape with large probability must therefore have the topography of a deep valley, or funnel}~\cite{Khoury:2019yoo,Khoury:2019ajl,Khoury:2021grg,Khoury:2022ish}.
This is akin to the smooth folding funnels of protein conformation landscapes~\cite{proteins1}, as sketched on the left panel in Fig.~\ref{funnel}. 

A similar narrative holds in deep learning. It has been argued that deep neural networks that generalize well have a loss function characterized by a smooth funnel~\cite{DDNlossfunnel} --- see right panel in Fig.~\ref{funnel}. 
Another instance is the ``big valley'' hypothesis in combinatorial optimization ({\it e.g.}, the search space of the traveling salesman problem), where it is conjectured that local optima are clustered around the central global optimum~\cite{TSP}.

\item {\bf Dominant decay channel:} In this approximation where~$T_{ij}$ is~$0$ or~$1$, the measure~\eqref{Puni} simply counts the number~$s_I$ of ancestor vacua
that can reach~$I$:
\be
\boxed{P(I) \sim s_I \,.}
\label{PIsI}
\ee
This is a key result of our analysis. It entails that the probability of occupying a vacuum is proportional to the number of other nodes that can access it through sequences of unsuppressed~($T_{ij} \simeq 1$), downward transitions. 
This is illustrated in Fig.~\ref{number anc fig} for a trivial example. 

\end{itemize}

Thus the problem of determining probabilities on vacua is reduced to a problem of directed percolation. To see this, consider a region of the landscape shown in Fig.~\ref{percolation sketch}, comprised of a number of transient dS vacua (blue nodes) and terminals (red nodes). (Recall that in the downward approximation terminals include AdS/Minkowski vacua, as well as dS vacua with only upward decay channels.) Each dS transient has exactly one directed edge emanating from it, corresponding to its dominant decay channel.

Suppose transients decay primarily to terminals, as sketched on the left panel in Fig.~\ref{percolation sketch}. In this case, nodes in the region will generically have~$s_I \sim {\cal O}(1)$, corresponding to relatively low probability. From a percolation perspective, the region breaks down into many small disconnected components, and is therefore subcritical. Suppose, on the other hand, that transients decay primarily to other transients, as shown on the right panel. This
corresponds to the emergence of a giant directed component, wherein the bottom nodes have~$s_I \gg 1$, and therefore high probability.

It is clear from these simple considerations that the uniform-in-time probabilities~\eqref{Puni} favor regions of the landscape that are close to the directed percolation phase transition~\cite{Odor:2002hk}.
In what follows we will make this precise by studying directed percolation on random graphs and Bethe lattices. 

\begin{figure}
\begin{center}
    \includegraphics[width=0.3\textwidth]{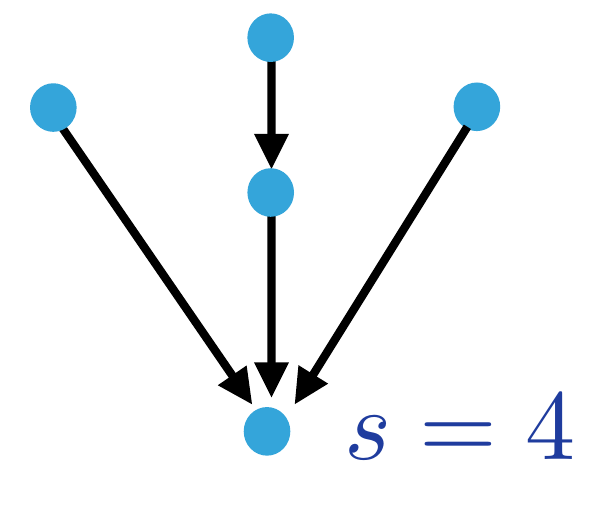}
  \caption{\label{number anc fig} In the downward and dominant decay channel approximations, the probability for a given node is proportional to the number of its ancestors. In this simple example, the bottom node has~$s = 4$ ancestors.}
\vspace{-5pt}  
\end{center}
\end{figure}

\section{Percolation on Directed Random Networks}
\label{sec:percolation}

To set up the problem, it is useful to review some essential notions of directed percolation~\cite{percolationtextbook,Essamreview}. Concretely, in this work we study two simplified approaches for directed percolation on the landscape. In the first framework, discussed in this section, we model the a fiducial region of the landscape as a directed random graph with given degree probability distribution. As a special case, the Poissonian degree distribution corresponds to the celebrated Erd\"os-R\'enyi graph~\cite{Erdos:1959:pmd}. In the second approach, discussed in Appendix~\ref{Bethe lattice sec}, we model the region as a regular lattice, specifically a Bethe lattice.

The directed percolation transition can be studied analytically for both Bethe lattices and (Erd\"os-R\'enyi) random graphs~\cite{PhysRevE.64.026118}. (In fact, they belong to the same universality class, as we will see.)
Our focus is on bond percolation, in which the percolation problem on either the Bethe lattice or the Erd\"os-R\'enyi random graph is defined by assigning a probability~$p$ that a given edge of the graph is ``open". While the frameworks considered are highly idealized, they allow us to draw important lessons about percolation phenomena on the landscape, which we believe apply more generally to realistic dynamics.

Although much of the analysis is already in the literature, we include it here for completeness. The reader mainly interested in the punchline can skip to~\eqref{PsPt ER}, which is the main result for our purposes. 
Our exposition primarily follows~\cite{PhysRevE.64.026118}, which considers random graphs with general degree distributions. For pedagogical purposes, we have also included in Appendix~\ref{perc undirected} a review of
percolation on undirected random graphs. Many of the results for the undirected case can be easily generalized to directed random graphs.

\subsection{Directed random graphs}

A directed random graph is specified by a joint in-degree and out-degree probability distribution:
\be
p_{jk} = \text{probability of randomly-chosen node having in-degree~$j$ and~out-degree $k$}\,.
\ee
It is useful to work in terms of its moment generating function,
\be
{\cal G}(x,y)  = \sum_{j,k = 0}^\infty p_{jk} x^j y^k\,.
\label{Gxy}
\ee
Since the distribution is normalized, we have~${\cal G}(1,1) = \sum_{j,k} p_{jk} = 1$. Its partial derivatives give in- and-out degree moments of the distribution.
For instance, the average in- and out-degrees are given by
\be
z^{\rm in} = \sum_{j,k} j p_{jk}  = \left.\frac{\partial {\cal G}(x,y)}{\partial x} \right\vert_{x = y = 1}\,; \qquad  z^{\rm out} =  \sum_{j,k} k p_{jk}  = \left.\frac{\partial {\cal G}(x,y)}{\partial y} \right\vert_{x = y = 1}\,.
\label{zinout}
\ee
Since every link leaving a node terminates at another node, the average in- and out-degrees must be equal:
\be
z^{\rm in} = z^{\rm out} \equiv z \,.
\label{z cond}
\ee
From~$p_{jk}$, we can derive the (marginalized) in- and out-degree distributions of a randomly-chosen vertex, with generating functions
\bea
\nonumber
F_0(x) &=& {\cal G}(x,1) = \sum_{j,k} p_{jk} x^j; \\
G_0(y) &=& {\cal G}(1,y) = \sum_{j,k} p_{jk} y^k \,\,.
\label{G0F0}
\eea
In particular, we have~$F_0'(1) =G_0'(1) =  z$.

\begin{figure}
\begin{center}
    \includegraphics[width=0.7\textwidth]{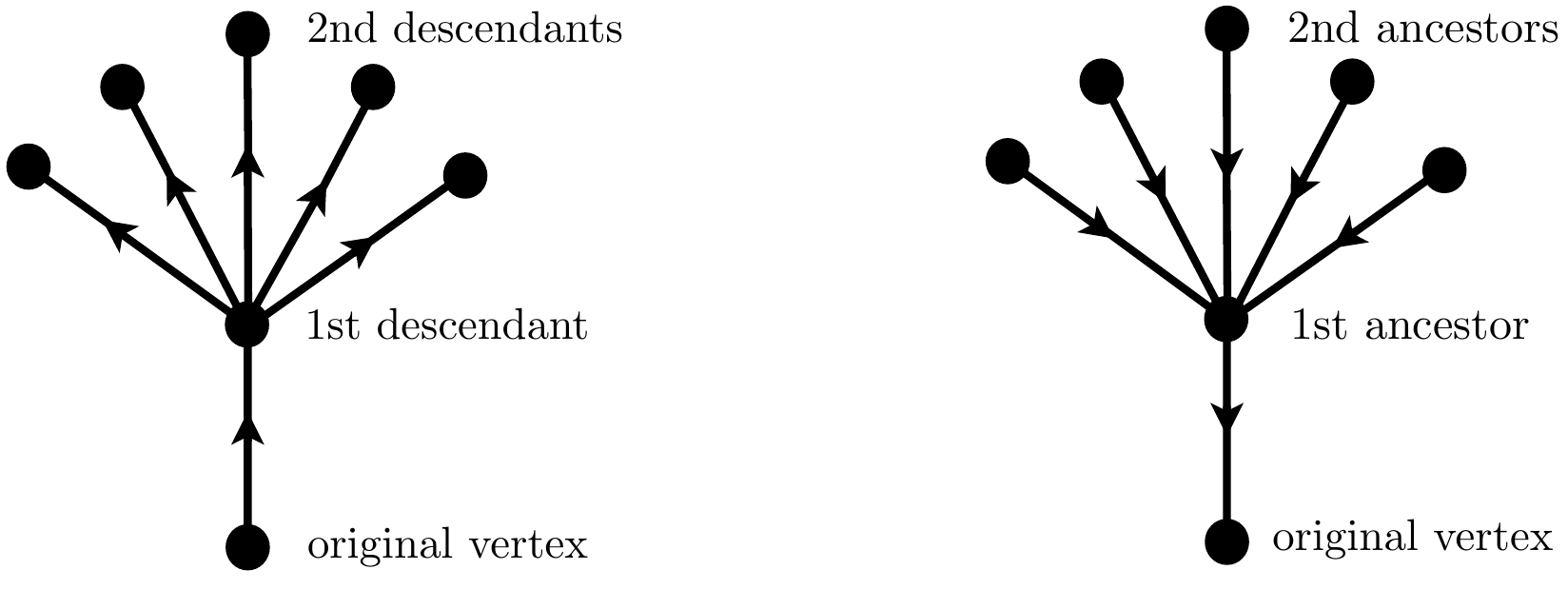}
  \caption{\label{1stneighbor directed} {\it Left:} 1st and 2nd-generation descendants of an original vertex. {\it Right:} 1st and 2nd-generation ancestors of that vertex.}
\vspace{-5pt}  
\end{center}
\end{figure}

Now, suppose we start from a randomly-chosen vertex, and follow each of its outgoing links to reach its 1st-generation descendants (children), as shown on the left panel of Fig.~\ref{1stneighbor directed}.
(In the directed case, ignoring loops, it is natural to distinguish the neighbors of a node as descendants and ancestors.\footnote{The terminology is quite apt close to the percolation threshold,
given the equivalence with branching processes. See, {\it e.g.},~\cite{bollobasriordan}.}) Let us denote by~$q_k^{\rm out}$ the out-degree distribution of a 1st descendant. Since we are~$j$ times more likely
to arrive at a vertex with in-degree~$j$ than a vertex of degree~$1$, we have
\be
q_k^{\rm out} = \frac{1}{z} \sum_{j} j p_{jk}\,,
\label{qk out}
\ee
which is correctly normalized. The corresponding generating function is given by
\be
G_1(y) =  \frac{1}{z}  \sum_{j,k} j p_{jk} y^k  = \frac{1}{z}  \left.\frac{\partial {\cal G}(x,y)}{\partial x} \right\vert_{x = 1} \,.
\label{G1y}
\ee
If the original vertex has out-degree~$k$, then the number of 2nd descendants is generated by~$\big(G_1(y)\big)^k$. (This ignores loops, since their density is~$1/N$-suppressed close to the percolation threshold for large~$N$, as argued in Appendix~\ref{perc undirected}.) Therefore, the number of 2nd descendants is generated by~$\sum_{j,k} p_{jk} \left(G_1(y)\right)^k = G_0\big(G_1(y)\big)$. For instance, using~\eqref{zinout},~\eqref{G0F0} and~\eqref{G1y}, the average number of 2nd descendants is
\be
z_2 = z G_1'(1) = \left. \frac{\partial^2{\cal G}}{\partial x\partial y} \right\vert_{x = y = 1}   \,.
\label{z2 directed}
\ee

Similarly, suppose we once again start from a randomly-chosen vertex, but now follow each of its incoming links in the opposite direction to reach its 1st ancestors (parents). 
See right panel of Fig.~\ref{1stneighbor directed}. By similar reasoning, the in-degree distribution for a 1st ancestor is generated by
\be
F_1(x) = \frac{1}{z}  \left.\frac{\partial {\cal G}(x,y)}{\partial y} \right\vert_{y = 1} \,.
\label{F1}
\ee
The number of 2nd ancestors of the original vertex is generated by~$\sum_{j,k} p_{jk} \left(F_1(x)\right)^j = F_0\big(F_1(x)\big)$.
For instance, the average number of 2nd ancestors is~$z F_1'(1) =  \left. \frac{\partial^2{\cal G}}{\partial x\partial y} \right\vert_{x = y = 1}$. This is of course identical to~\eqref{z2 directed},
given that it has~$x \leftrightarrow y$ symmetry.

\subsection{Factorized example} 
\label{fac dis sec}

\noindent An important example is the case where the in- and out-degree distributions are independent,
\be
p_{jk} = p^{\rm in}_j p^{\rm out}_k\,.
\label{pin pout factor}
\ee
This includes, as a particular case, Erd\"os-R\'enyi random graphs~\cite{Erdos:1959:pmd}. In a directed Erd\"os-R\'enyi graph with~$N$ vertices, any two distinct
vertices can be connected with a directed edge with probability~$p$. Therefore both~$p^{\rm in}_j$ and~$p^{\rm out}_k$ are given by a binomial distribution:
\be
p^{\rm out}_k = {N\choose k} p^k  (1-p)^{N-k} \simeq z^k \frac{{\rm e}^{-z}}{k!}\,,
\label{Poisson} 
\ee
and similarly for~$p^{\rm in}_j$. The last step follows from taking the limit $N\rightarrow \infty$, keeping the average degree~$z = p (N-1)$ fixed, to obtain a Poisson distribution.

The generating function~\eqref{Gxy} also factorizes,
\be
{\cal G}(x,y)  = G_0(y) F_0(x) \,,
\label{Gxy factor}
\ee
with~$G_0(y) = {\rm e}^{z(y-1)}$ and~$F_0(x) = {\rm e}^{z(x-1)}$. An immediate consequence of~\eqref{Gxy factor} is that~$G_1(y) = G_0(y)$, {\it i.e.}, the out-degree distribution for a 1st descendant is the same as that of the original vertex. Similarly,~$F_1(x) = F_0(x)$. It follows that the average number of 2nd descendants (or ancestors), given by~\eqref{z2 directed}, satisfies
\be
z_2 = z G_0'(1) = z^2\,.
\label{z fac}
\ee
This holds for arbitrary factorized distribution~\eqref{pin pout factor}, including the Erd\"os-R\'enyi case.


\begin{figure}
\begin{center}
    \includegraphics[width=1.0\textwidth]{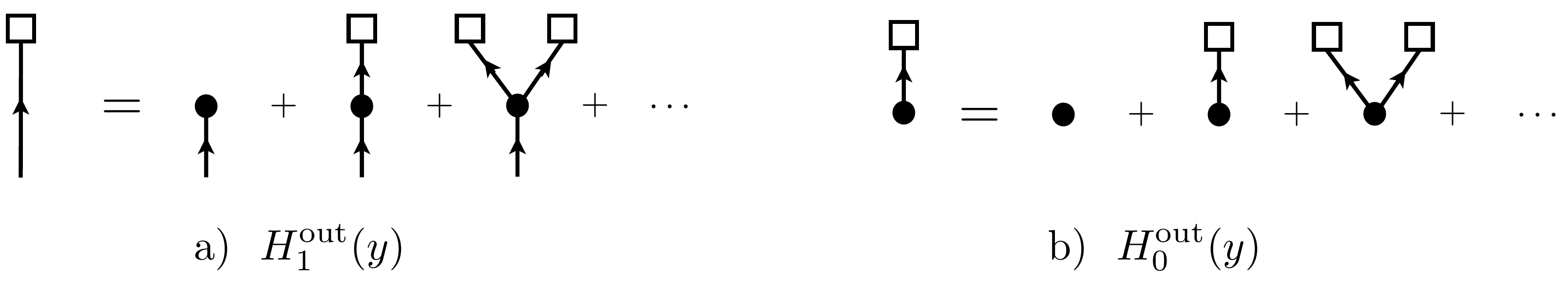}
  \caption{\label{H1 out tree} a) Tree-like structure satisfied by the generating function~$H_1^{\rm out}$ for the number of descendants following a randomly-chosen edge; b) Same structure, but for the generating function~$H_0^{\rm out}$ starting from a randomly-chosen vertex.}
\vspace{-5pt}  
\end{center}
\end{figure}

\subsection{In- and out-component size distribution and percolation phase transition}
\label{in out comp size sec}

\noindent Going back to the general case, the generating functions defined earlier allow us to study the size distribution of connected components. In the directed case, we must distinguish between the out-component, comprised of all descendants of a given vertex, and the in-component, comprised of all ancestors of a given vertex. 

Let us focus for concreteness on the out-component, and define
\be
H_1^{\rm out}(y) =  \text{Gen. fcn for number of descendants following a randomly-chosen edge.}
\ee
Ignoring loops, this generating function satisfies the tree-like consistency condition depicted in Fig.~\ref{H1 out tree}a).  From the definition~\eqref{qk out} for~$q_k^{\rm out}$,
the tree-like structure implies the consistency condition 
\be
H_1^{\rm out}(y) =  y \sum_{k = 0}^\infty q_k^{\rm out} \big(H_1^{\rm out} (y)\big)^k = yG_1\left(H_1^{\rm out}(y)\right)\,,
\label{H1 out implicit}
\ee
where the last step follows from~$q_k^{\rm out}$ being generated by~$G_1(y)$. Note that the factor of~$y$ means that the chosen node is included in the counting.

Similarly, we define
\be
H_0^{\rm out}(y) =  \text{Gen. fcn for number of descendants of a randomly-chosen vertex.}
\ee
This generating function satisfies the tree-like consistency condition depicted in Fig.~\ref{H1 out tree}b). Using the fact that~$G_0(y)$ is the generating function for the out-degree distribution
of a randomly-chosen vertex, the consistency condition in this case sums up to
\be
H_0^{\rm out}(y) = yG_0\left(H_1^{\rm out}(y)\right)\,.
\label{H0 out implicit}
\ee
By definition,~$H_0^{\rm out}$ describes finite components, {\it i.e.}, it excludes the giant out-component. As long we work below the percolation threshold, such that there is no infinite cluster, then~$H_0^{\rm out}(1) = 1$.
Above the percolation threshold,~$H_0^{\rm out}(1)$ gives the fraction of the vertices that do not belong to the giant component. See Appendix~\ref{perc undirected} for more details in the undirected case.

Note that the conditional probability $P^{\rm in}(s|k)$ for an in-cluster having size~$s$, given~$k$ in-degree of a vertex, is generated by
\begin{align}
    {\cal P}^{\rm in}(x|k) = \sum_s x^s P^{\rm in}(s|k)\,.
\end{align}
It should be related to the generating function for the number of ancestors~$x^{-1}H_0^{\rm in}(x) = F_0\left(H_1^{\rm in}(x)\right) $ via the definition 
\begin{align}
 F_0\left(H_1^{\rm in}(x)\right)= x^{-1}H_0^{\rm in}(x) = \sum_{k,\ell} p_{k\ell}{\cal P}^{\rm in}(x|k)\,.
\end{align}
Therefore by comparison we have 
\begin{align}
  {\cal P}^{\rm in}(x|k) =  \left(H_1^{\rm in}(x)\right)^k\,.
\end{align}
A similar derivation applies to the out-cluster conditional probability.

The algorithm for determining~$H_0^{\rm out}$ and~$H_1^{\rm out}$ is then the following. Given a joint degree distribution~$p_{jk}$, with generating function~${\cal G}(x,y)$, we
can determine~$G_0(y)$ and~$G_1(y)$ using~\eqref{G0F0} and~\eqref{G1y}, respectively. Then, the implicit equation~\eqref{H1 out implicit} can be solved to obtain~$H_1^{\rm out}(y)$, and the result is substituted
into~\eqref{H0 out implicit} to obtain~$H_0^{\rm out}(y)$.


For general random graphs, it is often difficult in practice to solve~\eqref{H1 out implicit} analytically. It is, however, straightforward to calculate the moments of the size distribution, in particular 
the {\it average size of (finite) connected components}. For instance, consider the average size~$S_{\rm out}(z)$ of the out-component reached from a random vertex.
For simplicity we work below the percolation threshold, such that there is no giant component, and~$H_0^{\rm out}(1) = H_1^{\rm out}(1) = 1$. Using~\eqref{H0 out implicit}, we have
\be
S_{\rm out}(z) = H_0^{{\rm out}\,'}(1) = 1 + G_0'(1) H_1^{{\rm out}\,'}(1)\,.
\label{mean s out}
\ee
On the other hand, from~\eqref{H1 out implicit} we have~$H_1^{{\rm out}\,'}(1) = 1 + G_1'(1) H_1^{{\rm out}\,'}(1)$, which implies~$H_1^{{\rm out}\,'}(1) = \frac{1}{1 - G_1'(1)}$.
Thus,
\be
S_{\rm out}(z)  = 1 + \frac{G_0'(1)}{1 - G_1'(1)} = 1 + \frac{z^2}{z - z_2}\,,
\ee
where we have used~\eqref{z2 directed}. Therefore a giant out-component emerges when
\be
z_2 = z \qquad (\text{directed percolation})\,.
\label{perc cond directed}
\ee

In exactly the same fashion, we can define generating functions~$H_1^{\rm in}(x)$ and~$H_0^{\rm in}(x)$ for the in-component size, obtain respectively by following a randomly-chosen
edge and starting from a randomly-chosen vertex. In doing so, we follow each incoming link in the opposite direction. These generating functions satisfy the implicit relations
\be
H_1^{\rm in}(x) = x F_1\left(H_1^{\rm in}(x)\right)\,;\qquad H_0^{\rm in}(x) = x F_0\left(H_1^{\rm in}(x)\right) \,.
\ee
Following identical steps as before, it is easy to derive the average size of the in-component:
\be
S_{\rm in}(z) = 1 + \frac{F_0'(1)}{1 - F_1'(1)} = 1 + \frac{z^2}{z - z_2}\,.
\ee
Therefore a giant in-component emerges when~$z_2 = z$, which is the same as~\eqref{perc cond directed}. 
In other words, the giant in- and out-components emerge simultaneously.\footnote{We should briefly comment on the meaning of the giant component in the directed case. In general directed graphs, there are three different types of giant components:~1) a strongly connected component, in which every vertex can reach every other vertex;~2) a component that contains vertices reachable from~1) but that cannot reach~1); and~3) a component comprised of vertices that can reach~1) but are not reachable from~1). These can be visualized as a ``bow-tie" diagram~\cite{bowtie}. Slightly above the percolation threshold, only~1) and~2) are present, since loops are suppressed. In directed acyclic graphs, the strongly connected component never arises, as 
loops are forbidden altogether.}

Equation~\eqref{perc cond directed} holds for any degree distribution~$p_{jk}$, with one important assumption --- the distribution should have finite mean and variance, {\it i.e.}, finite~$z$ and~$z_2$.
All degree distributions with this property belong to the Erd\"os-R\'enyi percolation universality class. Those that do not, for instance because their degree variance diverges, belong to different universality classes. 
An important example of the latter are scale-free random graphs, discussed in Sec.~\ref{scale-free}. In the particular case of factorizable distributions, we can combine~\eqref{z fac} and~\eqref{perc cond directed}
to recover the classic result that percolation occurs when
\be
z_{\rm c} = 1\,.
\ee
In particular, for the Poisson distribution~\eqref{Poisson} of Erd\"os-R\'enyi graphs, the critical probability is~$p_{\rm c} \simeq \frac{1}{N}$.

\subsection{Tail component size distributions}
\label{tail comp size directed}

Of prime importance for our analysis is the tail of the in- and out-component size distributions, defined as
\bea
\nonumber
P_s &=& \text{probability that a randomly-chosen vertex has~$s$ ancestors}\,;\\
P_t &=& \text{probability that a randomly-chosen vertex has~$t$ descendants}\,.
\label{PsPt def}
\eea
Their generating functions are respectively~$H_0^{\rm in}(x)$ and~$H_0^{\rm out}(y)$. Close to the percolation threshold, the tail of these distributions takes the form~\cite{PhysRevE.64.026118}
\be
P_s \sim s^{-\tau} {\rm e}^{-s/s_{\rm max}}\,; \qquad (s\gg 1)\,,
\label{Ps tail}
\ee
and similarly for~$P_t$. As usual the correlation length~$s_{\rm max}$ diverges as we approach the phase transition, {\it e.g.},~$s_{\rm max} \sim \left\vert p-p_{\rm c}\right\vert^{-1/2}$ for Erd\"os-R\'enyi graphs. 
By studying the behavior of the generating functions near the percolation threshold, one can show that the critical exponent takes the value~\cite{PhysRevE.64.026118}
\be
\tau = \frac{3}{2}\,.
\label{alpha def}
\ee
(For completeness, we provide a brief proof of this result in Appendix~\ref{critical exp}.) In other words, at criticality the distributions have the universal power-law tail
\be
\boxed{P_s \sim s^{-3/2}\,;\qquad P_t \sim t^{-3/2} \qquad (s,t \gg 1)} \,.
\label{PsPt ER}
\ee
Equation~\eqref{PsPt ER} is a key result for us, as it will play an important role in deducing the probability distribution for the CC in~Sec.~\ref{CC distribution}.

The typical size~$s_\star$ of the giant component can be estimated as the value at which the probability is~$1/N$. In other words,~$P_s(s_\star) \sim s_\star^{-3/2} \sim 1/N$,
and similarly for~$P_t$. Thus, at criticality the giant in- and out-components are both of size
\be
s_\star \sim N^{2/3}\,.
\label{sstarER}
\ee
All other components have size~${\cal O}(\log N)$. This holds for all random graphs with finite~$z$ and~$z_2$.


\subsection{Scale-free networks}
\label{scale-free}

A class of networks that has attracted much attention in the last two decades are scale-free networks~\cite{SF1,SFreview}. Empirically, many real-world networks exhibit this property, including the World Wide Web, social/collaboration networks and metabolic networks. Scale-free graphs include ``hubs" --- nodes connected to a very large number of other nodes. We briefly review the percolation structure on such networks, focusing for simplicity on undirected graphs, and refer the reader to~\cite{scalefreeexponents,scalefreepercolation} for details.

Scale-free networks are characterized by a degree distribution with power-law tail
\be
p_k \sim k^{-\gamma}\,.
\label{SF dis}
\ee
We require~$\gamma > 2$ in order for the distribution to be normalizable and have finite mean. If~$\gamma > 3$, such that the variance is also finite, then the percolation
structure belongs to the Erd\"os-R\'enyi universality class. So the interesting regime is
\be
2 < \gamma < 3\,.
\label{gamma 2 3}
\ee
A concrete example is 
\be
p_k =  \left\{\begin{array}{cl}
1 - z \frac{\zeta(\gamma)}{\zeta(\gamma - 1)} & ~~~k = 0  \,; \vspace{0.3cm} \\
\frac{z}{\zeta(\gamma - 1)} \, k^{-\gamma}  & ~~~k\geq 1 \,,
\end{array}\right.
\ee
where~$\zeta$ is the Riemann zeta function. The fraction of nodes with~$k = 0$ ensures that the distribution is normalized and has mean degree~$z$. 
Like Erd\"os-R\'enyi graphs, percolation occurs when~$z_{\rm c} = 1$~\cite{SF1,SFreview}.

At percolation criticality, the distribution of component sizes also exhibits a power-law tail, but with a different critical exponent:
\be
P_s \sim s^{-\frac{\gamma}{\gamma-1}}\,.
\label{Ps tail SF}
\ee
(Notice that the power matches~\eqref{alpha def} as~$\gamma\rightarrow 3$.) Thus each value of~$\gamma$ in the range~\eqref{gamma 2 3} defines its own universality class, comprised of all degree distributions with a scale-free tail with this particular power. The typical size of the giant component is estimated as before by setting~$P_s(s_\star) \sim s_\star^{-\frac{\gamma}{\gamma-1}} \sim 1/N$, which gives
\be
s_\star \sim N^{\frac{\gamma -1}{\gamma}}\,.
\label{sstar SF}
\ee

\section{Directed Percolation in Eternal Inflation}
\label{DP eternal infn}

Lacking detailed knowledge of the underlying string landscape, it is reasonable to model a fiducial landscape region as a random network.
With the quantitative results of the previous section at hand, let us briefly recap the approximations underlying the mapping to directed percolation.

\begin{enumerate}

\item The downward approximation, in which upward transitions are neglected to leading order. Strictly speaking, the downward approximation requires us to study directed acyclic graphs, {\it i.e.}, without directed loops.
However, as argued earlier, at low connectivity the density of cycles is suppressed by~$1/N$, hence directed random graphs offer a reasonable approximation.

\item The dominant decay channel approximation, in which the branching ratio~$T_{Ij}$ is either~0 or~1. This relies on semi-classical transition rates in field theory being exponentially staggered, and
therefore generically dominated by a single decay channel. 

\end{enumerate}

\noindent Let us stress that these approximations are made for convenience, to simplify the problem. It is in principle straightforward to generalize our analysis by relaxing them. For instance,
if a landscape region includes vacua that are nearly degenerate, such that the downward approximation is invalid, then the corresponding links would be bi-directed. The problem of directed percolation
with a finite fraction of bi-directed edges was studied in~\cite{BogunaSerrano}, where it was shown that bi-directed edges act as a catalyst for directed percolation. Similarly, if subdominant transitions are not completely negligible, 
such that the dominant decay approximation is invalid, then the corresponding network would be a random weighted graph~\cite{randomweightedgraph}. 

\subsection{Generating functions}

With these provisos in mind, consider a landscape region with~$N_{\rm dS}$ transient nodes (dS vacua) and~$N_{\rm AdS}$ terminal nodes. Although the latter also include dS vacua with
only upward decay channels, as well as Minkowski vacua, we use the collective ``AdS" subscript for simplicity. 

The moment generating function~\eqref{Gxy} can be written as
\be
{\cal G}(x,y)  = \frac{N_{\rm dS}}{N} {\cal G}^{\rm dS}(x,y) +  \frac{N_{\rm AdS}}{N}{\cal G}^{\rm AdS}(x,y) \,,
\label{Gxy dS AdS}
\ee
with~$N =N_{\rm dS} + N_{\rm AdS}$. Since terminals by definition have vanishing out-degree, we have
\be
{\cal G}^{\rm AdS}(x,y) = \sum_{k} p_{0 k}^{\rm AdS} x^j  = F_0^{\rm AdS}(x) \,,
\ee
where we have used~\eqref{G0F0}. Furthermore, since transients have out-degree~1 in the dominant decay channel approximation, as depicted in Fig.~\ref{branch ratio fig}, we should set~${\cal G}^{\rm dS}(x,y) = y \sum_{k} p_{1 k}^{\rm dS} x^j = y F_0^{\rm dS}(x)$. However, we will proceed more generally for now, and specialize to~$z_{\rm out}^{\rm dS} \simeq 1$ at the end of the calculation. 

The generating functions~\eqref{G0F0} for the marginalized in- and out-degree distributions of a randomly-chosen node are given by
\bea
\nonumber
G_0(y) &=&  \frac{N_{\rm dS}}{N} G_0^{\rm dS}(y)   +  \frac{N_{\rm AdS}}{N} \,; \\
F_0(x) &=& \frac{N_{\rm dS}}{N} F_0^{\rm dS}(x) +   \frac{N_{\rm AdS}}{N} F_0^{\rm AdS}(x)  \,.
\label{G0F0 dS AdS}
\eea
The condition~\eqref{z cond} for edge conservation gives
\be
z = \frac{N_{\rm dS}}{N} z_{\rm in}^{\rm dS} + \frac{N_{\rm AdS}}{N} z_{\rm in}^{\rm AdS} = \frac{N_{\rm dS}}{N} z_{\rm out}^{\rm dS}\,.
\label{z dS AdS}
\ee

Next, the out-degree distribution for a 1st descendant, given by~\eqref{G1y}, amounts to weighing by the number of edges:
\be
G_1(y) = \frac{1}{z_{\rm out}^{\rm dS}} \left(\frac{N_{\rm AdS}}{N_{\rm dS}} z_{\rm in}^{\rm AdS} + \left.\frac{\partial {\cal G}^{\rm dS}(x,y)}{\partial x} \right\vert_{x = 1}\right)  \,.
\label{G1y dS AdS}
\ee
Similarly, the in-degree distribution for a 1st ancestor, given by~\eqref{F1}, reduces to
\be
F_1(x) = \frac{1}{z_{\rm out}^{\rm dS}}  \left.\frac{\partial {\cal G}^{\rm dS}(x,y)}{\partial y} \right\vert_{y = 1}  \,.
\label{F1 dS AdS}
\ee
The number of 2nd ancestors of a given vertex is generated by~$\frac{N_{\rm dS}}{N} F_0^{\rm dS}\big(F_1(x)\big) +   \frac{N_{\rm AdS}}{N} F_0^{\rm AdS}\big(F_1(x)\big)$.
In particular, the average number of 2nd ancestors, which equals the average number of 2nd descendants, is
\be
z_2 = \frac{N_{\rm dS}}{N}  \left. \frac{\partial^2{\cal G}^{\rm dS}}{\partial x\partial y} \right\vert_{x = y = 1}   \,.
\label{z2 dS AdS}
\ee

\subsection{Percolation phase transition}

The derivation of the directed percolation phase transition given in Sec.~\ref{in out comp size sec} follows identically in the case of interest. For instance,
the generating functions~$H_1^{\rm out}(y)$ and~$H_0^{\rm out}(y)$ for the number of descendants satisfy the same implicit relations~\eqref{H1 out implicit} and~\eqref{H0 out implicit}:
\be
H_1^{\rm out}(y) = yG_1\left(H_1^{\rm out}(y)\right)\,;\qquad
H_0^{\rm out}(y) = yG_0\left(H_1^{\rm out}(y)\right)\,,
\ee
with~$G_0$ and~$G_1$ respectively given by~\eqref{G0F0 dS AdS} and~\eqref{G1y dS AdS}.

From~\eqref{perc cond directed}, the directed percolation phase transition occurs when~$z_2 = z$. Using~\eqref{z dS AdS} and~\eqref{z2 dS AdS},
this means
\be
\left. \frac{\partial^2{\cal G}^{\rm dS}}{\partial x\partial y} \right\vert_{x = y = 1} = z_{\rm in}^{\rm dS} + \frac{N_{\rm AdS}}{N_{\rm dS}} z_{\rm in}^{\rm AdS}   = z_{\rm out}^{\rm dS}  \qquad (\text{directed percolation}) \,.
\label{perc cond dS AdS}
\ee
As mentioned earlier, consistent with the dominant decay channel approximation we should set~${\cal G}^{\rm dS}(x,y) = y F_0^{\rm dS}(x)$, such that the out-degree of transient vacua is precisely~1.
To see how percolation works out, it is instructive to keep things slightly more general by assuming that transients have independent  in- and out-degree distributions:
\be
{\cal G}^{\rm dS}(x,y) = G_0^{\rm dS}(y)F_0^{\rm dS}(x)\,.
\label{dS fac}
\ee
In this case the percolation condition~\eqref{perc cond dS AdS} reduces to
\be
z_{\rm in}^{\rm dS} = 1\,.
\ee
Equivalently, from~\eqref{z dS AdS},
\be
\boxed{z_{\rm out}^{\rm dS} = 1 +  \frac{N_{\rm AdS}}{N_{\rm dS}} z_{\rm in}^{\rm AdS}} \,.
\label{zout dS perc}
\ee

Equation~\eqref{zout dS perc} is a key result of our analysis. From the point of view of dS vacua, the presence of terminals pushes the percolation threshold above unity, {\it i.e.},~$z_{\rm out}^{\rm dS} > 1$. This
makes sense intuitively, as absorbing nodes inhibit the emergence of a giant component. On the other hand, the dominant decay channel approximation tells us that~$z_{\rm out}^{\rm dS}\simeq 1$. Therefore, in order for a landscape region to be near percolation criticality, it must satisfy
\be
N_{\rm AdS} z_{\rm in}^{\rm AdS} \ll N_{\rm dS} \,.
\ee
This is the situation shown in the right panel of Fig.~\ref{percolation sketch}, wherein dS vacua decay primarily to other transients, and the region includes
a giant funnel of size
\be
s_\star \sim  \left\{\begin{array}{cl}
N_{\rm dS}^{2/3}  & \text{ER class} \,; \vspace{0.3cm} \\
N_{\rm dS}^{\frac{\gamma - 1}{\gamma}} & \text{scale-free,}~~~ 2 < \gamma < 3 \,.
\end{array}\right.
\ee
 In contrast, if a significant fraction of dS vacua decay into terminals, such that~$N_{\rm AdS} z_{\rm in}^{\rm AdS}$ is comparable to~$N_{\rm dS}$,
then the landscape region will be subcritical, as shown in the left panel of Fig.~\ref{percolation sketch}.

As argued in~\eqref{PIsI}, in the downward and dominant decay channel approximations, the probability to occupy a node is proportional to the number of its ancestors:~$P(I) \sim s_I$.
In other words, vacua with high occupation probability have large number of ancestors. The probability that a randomly-chosen vacuum has~$s$ ancestors is precisely given
by~$P_s$, defined in~\eqref{PsPt def}. For subcritical regions, the tail distribution of the component is exponentially cut off, as in~\eqref{Ps tail}, and therefore vacua in such
regions typically have~$s \sim {\cal O}(1)$. For near-critical regions, however,~$P_s$ displays a power-law tail, given by~$P_s \sim s^{-3/2}$ for the ER universality class,
and~$\sim s^{-\frac{\gamma}{\gamma-1}}$ with~$2 < \gamma < 3$ for the scale-free graphs (see~\eqref{Ps tail SF}). Correspondingly, near-critical regions include nodes whose number of ancestors
is of order the size of the giant component, {\it i.e.},~$s \sim s_\star$.

Thus we arrive at an important realization. To the extent that landscape regions can be modeled as random networks, as we have done, we
conclude that {\it vacua with the highest occupational probability reside in landscape regions that are close to directed percolation criticality.}
As usual, near the percolation phase transition, various observables assume power-law (scale-invariant) probability distributions, characterized by
universal critical exponents that are insensitive to the microscopic details of the system. As we are about to show, the critical exponent for~$P_s$
translates to a critical exponent for the CC distribution.

\section{Critical Exponent for the Cosmological Constant}
\label{CC distribution}

Let~$f_V(v)$ denote the underlying CC probability distribution function on the landscape, where~$v = \Lambda/M_{\rm Pl}^4$ is the dimensionless CC. In what follows we will keep~$f_V(v)$ completely general, except for one assumption made at the end, namely that the distribution is smooth as~$v\rightarrow 0^+$, such that
\be
F_V(v) \simeq f_V(0) v\,; \qquad \text{for}~~0 < v \ll 1\,,
\label{fv smooth}
\ee
where~$F_V(v)$ is the cumulative distribution function.\footnote{Some authors~\cite{Sumitomo:2012wa,Sumitomo:2012vx,Sumitomo:2012cf,Danielsson:2012by,Sumitomo:2013vla,Tye:2016jzi} have argued that the underlying distribution~$f_V(v)$ for the landscape diverges as~$v\rightarrow 0^+$. If this is the case, this would make our critical exponent for the CC even more negative.}

Our task is to derive a probability distribution~$P(v)$ that takes into account the measure factor from cosmological dynamics. For this purpose, we focus on landscape regions close to directed percolation criticality.
As argued above, such regions include vacua whose number of ancestors are of order the size of the giant component, {\it i.e.},~$s \sim s_\star$, and therefore have very high probability. Furthermore, since all but
one vacuum in the giant component (in any connected component, for that matter) is a terminal, we are justified in focusing on $v >0$ to deduce the CC distribution. (This is an obvious consequence of the the
downward and dominant decay channel approximations.) In other words, vacua in the giant component are overwhelmingly more likely to be dS vacua than AdS. For concreteness we focus on the Erd\"os-R\'enyi
universality class, and briefly discuss the generalization to scale-free graphs at the end.

Consider a vertex in such a region, and suppose that this vertex has~$s$ ancestors and~$t$ descendants. If the vertex in question has vacuum energy~$v$, then in the downward approximation its~$s$ ancestors all have larger vacuum energy, while its~$t$ descendants all have lower vacuum energy. In other words, the conditional CC probability distribution~$P(v|s,t)$ follows ordered statistics:
\be
P(v|s,t) = \frac{(s+t+1)!}{t! s!} f_V(v) \big(F_V(v)\big)^t \big( 1 - F_V(v)\big)^s\,.
\ee
It is convenient to use the cumulative distribution itself as the random variable,~$U \equiv F_V$, such that
\be
{\rm d} u = f_V(v) {\rm d}v\,.
\ee
Clearly~$f_U(u)$ is uniform over~$u \in [0,1]$. Indeed, the ordered statistics of~$U$ are simply those of the uniform distribution:
\be
P(u|s,t) =  \frac{(s+t+1)!}{t! s!} u^t (1-u)^s = \text{Beta}(t+1,s+1)(u)\,.
\ee
The desired CC probability distribution is obtained by marginalizing over~$s$ and~$t$,
\be
P(u) =  \sum_{s,t} f_S(s) f_T(t) \,\text{Beta}(t+1,s+1)(u) \,,
\label{Pu gen}
\ee
where we have used the fact that~$s$ and~$t$ are independent random variables, even for correlated degree distributions~\cite{BogunaSerrano}. 

The probability distribution~$f_S(s)$ is given by
\be
f_S(s) \sim s P_s \,,
\label{fS}
\ee
where~$P_s$ is the probability that a randomly-chosen vertex has~$s$ ancestors, defined in~\eqref{PsPt def}. The factor of~$s$ is the cosmological measure factor~\eqref{PIsI}, which encodes the fact that the probability to pick
a node is proportional to the number of ancestors. Since~$P_s \sim s^{-3/2}$ in the tail at criticality,~$f_S(s)\sim s^{-1/2}$ is not normalizable for~$N_{\rm dS} \rightarrow\infty$. For the realistic case of finite~$N_{\rm dS}$, however, it is regularized by
the giant component of size~$s_\star \sim N^{2/3}_{\rm dS}$. It follows that~$f_S(s)$ has the power-law behavior
\be
f_S(s) \simeq  \frac{1}{2\sqrt{s_\star} \sqrt{s}} \,;\qquad s_\star \geq s \gg 1 \,.
\label{fS tail}
\ee
Meanwhile,~$f_T(t)$ is just the probability distribution~$P_t$ that a randomly-chosen vertex has~$t$ descendants. All we will need is its power-law tail behavior~\eqref{PsPt ER}:
\be
f_T(t)  = P_t \sim  \frac{1}{t^{3/2}} \,;\qquad t \gg 1 \,.
\label{fT tail}
\ee

\subsection{Analytic approximation for~$P(u)$}

We proceed to evaluate~\eqref{Pu gen} analytically. We assume throughout that~$u\ll 1$, and this will be justified {\it a posteriori} since the resulting distribution will peak for~$u\ll 1$. 
First, let us suppose that the integral is dominated by the tail region~$s,t\gg 1$, such that~$f_S(s)$ and~$f_T(t)$ are given by~\eqref{fS tail} and~\eqref{fT tail}. In this regime, the Beta
distribution is well-approximated by a Gaussian, and the sums can be approximated as integrals:
\be
P(u) \sim \int \frac{{\rm d}s}{2\sqrt{s_\star} \sqrt{s}} \int \frac{{\rm d}t}{t^{3/2}} \frac{(s+t)^{3/2}}{\sqrt{2\pi st}} \exp\left[- \frac{s+t}{2st} \left(t - su\right)^2\right]\,.
\label{Pu 2}
\ee
The~$t$ integral can be evaluated using Laplace's method. The exponent is stationary for~$t_0 = su$. Consistency of the tail approximation requires~$t_0 \gg 1$, and in particular
\be
s_\star u\gg 1\,.
\label{s_*u gg 1}
\ee
Evaluating the~$t$ integral, we obtain
\be
P(u) \sim \frac{1}{2\sqrt{s_\star}} u^{-3/2} \int^{s_\star}_{1/u} \frac{{\rm d}s}{s}  = \frac{1}{2 \sqrt{s_\star}} u^{-3/2} \ln \big(s_\star u\big)\,; \qquad s_\star^{-1} \ll u \ll 1 \,.
\label{Pu large}
\ee
This peaks for small~$u$, as anticipated. 

To see that the distribution is well-behaved as~$u\rightarrow 0$, consider the regime~$s_\star u\ll 1$. In this case we can make the approximation~$(1-u)^s \simeq 1$ for all~$s$,
such that~\eqref{Pu gen} becomes
\be
P(u) \simeq \sum_{s,t} f_S(s) f_T(t)  \,\frac{(t+s+1)!}{s!\,t!} \,u^t\,.
\ee
To proceed, let us assume that~$s \gg t$, which can be justified {\it a posteriori}. Thus we obtain
\be
P(u) \simeq \sum_{s,t}  \,f_S(s) s   f_T(t)  \frac{(sv)^t}{t!}\,.
\ee
Since~$s u\ll 1$, clearly the sum peaks for~$t = 0$, in which case
\be
P(u) \simeq f_T(0) \sum_{s}  \,f_S(s) s \simeq \frac{1}{3} f_T(0) s_\star\,; \qquad  u \ll s_\star^{-1} \,.
\label{Pu small}
\ee
Thus, as claimed the distribution is smooth for~$u \rightarrow 0$, and reaches a maximum value of~$\sim s_\star$ at the origin.

\subsection{The CC distribution}

Equations~\eqref{Pu large} and~\eqref{Pu small} give the general probability distribution for the cumulative variable~$u = F_V(v)$ of a general distribution~$f_V(v)$. To translate to a probability distribution~$P(v)$ for~$v$ itself,
we now make the assumption~\eqref{fv smooth} that~$f_V(v)$ is smooth as~$v\rightarrow 0^+$, such that
\be
u \simeq f_V(0) v \,.
\ee
In the ``tail", defined by~$f_V^{-1} (0)s_\star^{-1} \ll  v  \ll \text{min}\left(f_V^{-1}(0),1\right)$,  the CC probability distribution is a power-law
\be
\boxed{P(v) \sim \frac{1}{2\sqrt{f_V(0) s_\star}} \, v^{-3/2} \ln \big(s_\star f_V(0) v\big) }  \,.
\label{CC final}
\ee
Thus the~$-3/2$ critical exponent for~$s$ and~$t$ translates to an identical critical exponent for~$v$. 
In particular, the~$95\%$ confidence interval for the CC is
\be
v\lesssim \frac{\log 20}{f_V(0) s_\star} \sim N^{-2/3}_{\rm dS} \,.
\ee
Therefore the cosmological measure favors small, positive vacuum energy. It can explain our observed CC~$v_{\rm obs} \sim 10^{-120}$ if our vacuum belongs to a funneled region of size~$N_{\rm dS}\sim 10^{240}$.

To be clear~$N_{\rm dS}$ is the number of dS (transient) vacua in a funnel region near directed percolation criticality, not the total number of dS vacua across the entire landscape.
Since the cosmological measure favors vacua with the largest number of ancestors, we are likely to inhabit the largest funnel region near percolation criticality, {\it i.e.}, the near-critical
region with largest~$N_{\rm dS}$. 

Unfortunately, it is unlikely that the Erd\"os-R\'enyi CC distribution derived above offers a viable solution to the CC problem. The reason is that it likely suffers from an empty universe problem.
With a Poisson distribution, a given node typically only has a few parents. Their CC distribution is also given by~\eqref{CC final}, hence they are themselves likely to have a small CC. Since the vacuum energy
of the parents sets an upper bound on the energy scale of the last period of slow-roll inflation, the generic outcome is an empty universe. A possible way out is if low-lying nodes have a large number of parents. This
requires many other nodes in the giant component to be orphans ({\it i.e.}, have vanishing in-degree), in order to maintain the criticality condition~$z_{\rm in}^{\rm dS} = 1$.

Scale-free graphs offer a natural realization of this scenario. As  discussed in Sec.~\ref{scale-free}, scale-free graphs include hubs, whose in-degree is of order the size of the
giant component. It is straightforward to repeat the analysis for the scale-free degree distribution. The resulting power-law in this case is
\be
P(v) \sim v^{-\frac{\gamma}{\gamma - 1}} \qquad (2 < \gamma < 3)\,,
\ee
which is the same critical exponent as~$P_s$ itself, {\it cf.}~\eqref{Ps tail SF}. Using~\eqref{sstar SF}, the~$95\%$ confidence interval is
\be
v \lesssim N_{\rm dS}^{- \frac{\gamma - 1}{\gamma}}\,.
\ee
We will comment further on the empty universe problem and scale-free graphs in the Conclusions.

\section{Conclusions}
\label{sec:concl}

The near-criticality of our universe may be the strongest empirical hint that we are part of a multiverse. A natural arena to realize this ensemble is the vast energy landscape of string theory, together with the dynamics of eternal inflation to instantiate in space-time the different vacua of the landscape. Making robust statistical predictions for physical observables in our own universe is unquestionably a task of fundamental importance in theoretical physics. Yet, how can we
ever hope to make progress towards this goal without a detailed understanding of the string landscape?

Fortunately, despite all of its conceptual pitfalls, eternal inflation boils down to a random walk on the network of vacua. Technically this is an absorbing Markov process,
because of terminal (AdS/Minkowski) vacua which act as sinks. Hence the dynamics are inherently non-equilibrium. The Markov process leads to a natural definition of probabilities
as occupational probabilities for the random walk. 

%

In this paper we showed how the Markov process governing vacuum dynamics can be mapped naturally to a problem of directed percolation on the network of vacua.
The mapping relies on two very general and well-justified approximations for transition rates: 1.~the downward approximation, which neglects ``upward" transitions, as these
are generally exponentially suppressed; 2.~the dominant decay channel approximation, which capitalizes on the fact that tunneling rates are exponentially staggered. 

With these simplifying assumptions, we argued that the uniform-in-time probabilities reduce a simple and intuitive observable in directed graphs. Namely, the probability to occupy
a particular node is proportional to the number of its ancestors, {\it i.e.}, how many other nodes can reach it through a sequence of directed (downward) transitions. Thus the probabilities
favor vacua with a large basin of ancestors, lying at the bottom of a deep funnel. Funneled landscape topography appears to be a common solution to optimization on complex energy
landscapes, including protein folding~\cite{proteins1}, atomic clusters~\cite{Doye2002,Doye2004,Doye2005}, deep learning~\cite{DDNlossfunnel}, and combinatorial optimization~\cite{TSP}. 

Lacking detailed knowledge of the string landscape, we modeled the network of vacua as random graphs with arbitrary degree distributions, including Erd\"os-R\'enyi and scale-free graphs. As a complementary
approach, we also modeled regions of the landscape as regular lattices, specifically Bethe lattices. Despite representing extreme opposites of graph regularity, Bethe lattices and Erd\"os-R\'enyi belong to the same
percolation universality class. Thus one may hope that the lessons drawn from studying percolation in these simplified, idealized setups carry over to more realistic landscapes.

The most important result of our analysis is that the uniform-in-time probabilities favor regions of the landscape poised at the directed percolation phase transition. In other words, 
our vacuum most likely resides within a network of vacua tuned at directed percolation criticality. As usual, the predictive power of criticality lies in universality. This raises the tantalizing
prospect of deriving statistical predictions for physical observables that are insensitive to the details of the underlying landscape. More broadly, it suggests a deep and powerful relation
between phase transitions in landscape dynamics and the inferred near-criticality of our universe. 

To illustrate this point, we derived a probability distribution for the CC. At percolation criticality, the probability distributions for the number of ancestors and descendants of a
given node both display power-law tails, with certain critical exponents. Assuming only that the underlying CC distribution function is smooth near the origin, we derived
probability distributions for the CC that are also power-law, with critical exponents that are determined by the universality class (Erd\"os-R\'enyi or scale-free). These distributions favor small, positive
CC, and can account for the observed CC if our vacuum belongs to a large enough region. In fact, since the measure favors vacua with the largest number of ancestors, we are likely to inhabit the
largest funnel region near percolation criticality.

There are many future directions of inquiry worth pursuing. Let us mention two concrete follow-ups:

\begin{itemize}  
 
\item It is possible to derive probability distributions for other physical observables. To give one example, consider a node of in-degree~$k$. The joint probability distribution for its~$k$ parents to have~$s_1,s_2,\ldots,s_k$ ancestors,~$P(s_1,\ldots,s_k |k)$,
translates to a joint distribution for the potential energy of its parents, given by~$P(v_1,\ldots,v_k |k)$. This statistics informs us on the energy scale of the last period of slow-roll inflation, which of course has immediate bearing on the observational prospects of detecting primordial gravitational waves. This also has immediate bearing on the potential empty universe problem discussed in Sec.~\ref{CC distribution}. For instance, one could condition on sufficiently large~$k$, {\it i.e.}, large number of parents.

\item The connection with other complex energy landscapes deserves further exploration. A remarkable aspect of protein folding networks is that they are scale-free, with the native state acting as a hub with very large degree~\cite{Rao_Caflisch_2004}. Protein folding networks also display the small-world property, and are hierarchical. Similar properties are found in atomic clusters with Leonard-Jones interactions~\cite{Doye2002,Doye2004,Doye2005} --- their funnel topography is hierarchical (funnels nested within larger funnels), and the degree distribution is scale-free. It may be that such properties are generic to optimization on complex energy landscapes. It will be fascinating to explore their implications in the context of landscape dynamics. 

\end{itemize}

\vspace{.4cm}
\noindent
{\bf Acknowledgements:} We thank Giorgos Gounaris, James Halverson, Eleni Katifori, Cody Long, Minsu Park, Anushrut Sharma and Nathaniel Watkins for helpful discussions. This work is supported in part by the US Department of Energy (HEP) Award DE-SC0013528.

\begin{figure}
\begin{center}
    \includegraphics[width=0.3\textwidth]{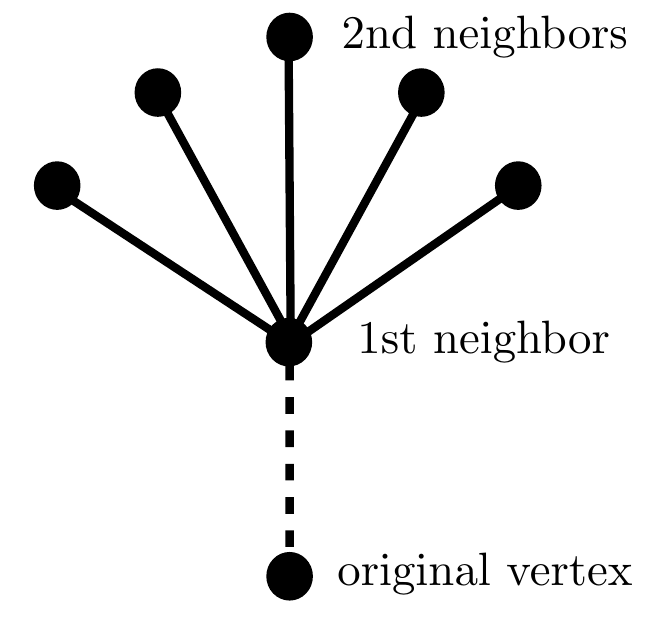}
  \caption{\label{1stneighbor}Starting from a randomly-chosen (``original") vertex, we follow of its links (dashed line) to a 1st neighbor. The probability that this 1st neighbor has~$k$ edges excluding the one we followed is~$q_k$, with generating function~$G_1(x)$ given by~\eqref{G1}.}
\vspace{-5pt}  
\end{center}
\end{figure}

\begin{appendices}

\section{Percolation on Undirected Random Graphs}
\label{perc undirected}

In this Appendix, we give a review of percolation on undirected random graphs, for completeness. An undirected random network is defined by specifying a degree probability distribution:
\be
p_k = \text{probability of a randomly-chosen node having degree}~k\,.
\ee
The moment generating function is
\be
G_0(x)  = \sum_{k = 0}^\infty p_k x^k\,,
\label{G0}
\ee
with normalization condition~$G_0(1) = \sum_{k= 0}^\infty p_k = 1$.
Its derivatives give as usual the moments of the distribution, such as the average degree:
\be
z \equiv \langle k \rangle  = \sum_{k = 0}^\infty k p_k  = G_0'(1)\,.
\label{z}
\ee

The next important quantity is the degree distribution of a vertex reached by following a randomly-chosen edge. This distribution is not simply~$p_k$, since we are~$k$ times more likely to arrive at a vertex with degree~$k$ as we are at a vertex of degree~$1$. Therefore this distribution is proportional to~$k p_k$. Now, suppose we start from a randomly-chosen vertex, and follow each of its links to reach the 1st neighbors. We are interested in the ``excess degree" of the 1st neighbors, which excludes the edge we arrived along.
Let 
\be
q_k = \text{probability that 1st neighbor has excess degree~$k$}\,.
\ee
This is shown in Fig.~\ref{1stneighbor}. This distribution is related to~$p_k$ by~$q_k = \frac{(k+1)p_{k +1}}{\sum_k k p_k} = \frac{(k+1)p_{k+1}}{z}$, 
and its generating function is given by
\be
G_1(x) \equiv \sum_{k=0}^\infty q_k x^k = \frac{G_0'(x)}{z} \,.
\label{G1}
\ee

Now, if the original vertex has degree~$k$, then the number of second-nearest neighbors is generated by~$\big(G_1(x)\big)^k$. This ignores loops, which are negligible in the large~$N$ limit below the percolation threshold, as argued below. It follows that number of 2nd neighbors is generated by~$\sum_{k=0}^\infty p_k \left(G_1(x)\right)^k = G_0\big(G_1(x)\big)$. 
For instance, the average number of 2nd neighbors is
\be
z_2 = \left.\frac{{\rm d}}{{\rm d}x} G_0\big(G_1(x)\big)\right\vert_{x = 1} = G_0''(1)\,,
\label{z2}
\ee
where we have used~\eqref{z} and~$G_1(1) = 1$.

\subsection{Erd\"os-R\'enyi example} 

\noindent An important example is the Erd\"os-R\'enyi graph~\cite{Erdos:1959:pmd}. An undirected Erd\"os-R\'enyi graph is a random graph with~$N$ vertices, in which an edge between any two distinct vertices has a probability~$p$ of being included. Therefore the probability distribution~$p_k$ is binomial and tends to the Poisson distribution for large~$N$, as given by~\eqref{Poisson}. The generating functions~\eqref{G0} and~\eqref{G1} in this case are given by
\be
G_0(x) = G_1(x) = {\rm e}^{z(x-1)}\,.
\label{G0G1ER}
\ee
The average number of 2nd neighbors~\eqref{z2} is~$p^2(N-1)(N-2)\simeq p^2N^2$, and therefore
\be
z_2 \simeq z^2 \qquad (\text{Poisson})\,.
\label{z2z Poisson}
\ee

To see that the probability of having loops is suppressed by~$\frac{1}{N}$ below or close to the percolation threshold, consider the probability of having two 1st neighbors connected to each other. For a random vertex with degree~$k$ there are~$\frac{k(k-1)}{2}$ possible edges between any pair of its neighbors. The chance of having at least one loop is~$1-(1-p)^{\frac{k(k-1)}{2}}$. In Erd\"os-R\'enyi graphs,~$p \simeq \frac{z }{N}$, thus the chance of having triangular loops is~$\simeq \frac{z k(k-1) }{2N}\sim \frac{1}{N}$ in the large~$N$ limit.


\begin{figure}
\begin{center}
    \includegraphics[width=1.0\textwidth]{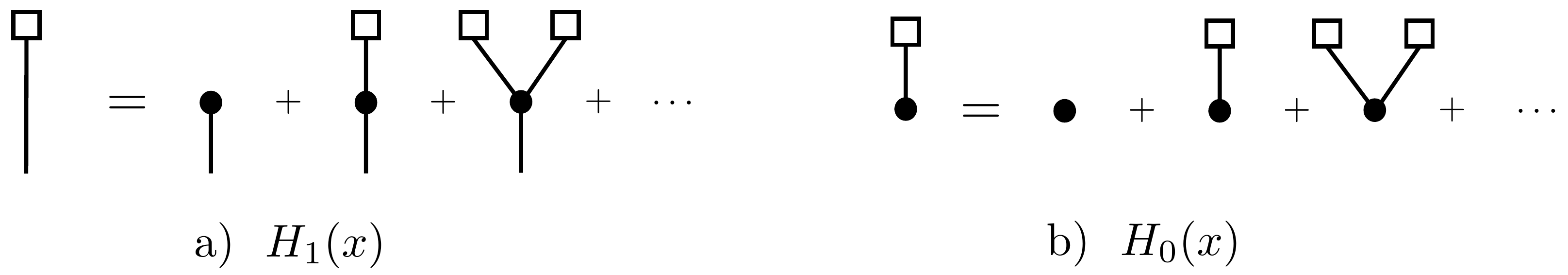}
  \caption{\label{H1tree} a) Tree-like structure satisfied by the generating function~$H_1$ for the distribution of component sizes starting from a randomly-chosen edge; b) Same structure, but for the generating function~$H_0$ for the
  component sizes starting from a randomly-chosen vertex.}
\vspace{-5pt}  
\end{center}
\end{figure}

\subsection{Component size distribution}

With these tools at hand, we next consider the size distribution of connected components. For this purpose we work below (or close to) the percolation phase transition.
We define
\be
H_1(x) =  \text{Gen. fcn for size of components reached by following a randomly-chosen {\it edge}.}
\label{H1 def}
\ee
Ignoring loops, as they are suppressed by~$\frac{1}{N}$ close to the percolation threshold, we have the tree-like structure depicted in Fig.~\ref{H1tree}a). From the definition of~$q_k$, the tree-like structure implies the consistency condition 
\be
H_1(x) = x \sum_{k = 0}^\infty q_k \big(H_1(x)\big)^k  = xG_1\big(H_1(x)\big)\,,
\label{H1 implicit}
\ee
where the last step follows from~$q_k$ being generated by~$G_1$.
Similarly, we define the distribution of (finite) component sizes reached starting from a randomly-chosen {\it vertex}:
\be
P_s = \text{probability that (finite) component reached by following a randomly-chosen {\it vertex} has size}~s\,.
\label{Ps def}
\ee
We denote by~$H_0(x)$ the corresponding generating function:~$H_0(x) = \sum_{s=0}^\infty P_s x^s$.
It satisfies the consistency relation, depicted in Fig.~\ref{H1tree}b),
\be
H_0(x) = x \sum_{k = 0}^\infty p_k \big(H_1(x)\big)^k = xG_0\big(H_1(x)\big) \,,
\label{H0 implicit}
\ee
where the last step follows from~$p_k$ being generated by~$G_0$. By definition,~$H_0$ describes finite components, {\it i.e.}, it excludes the giant component. As long we are below the percolation threshold, such that there is no infinite cluster, then~$H_0(1) = 1$. Above the percolation threshold,~$H_0(1)$ gives the fraction of the vertices that do not belong to the giant component. To be precise, let~$P(z)$ denote the fraction of vertices belonging to the
giant component: 
\be
P(z) = 1 - H_0(1)\,.
\label{P(z) def}
\ee

\subsection{Percolation phase transition}
\label{perc PT sec}

For general random graphs, it is often difficult in practice to solve~\eqref{H1 implicit} analytically, hence we must content ourselves with computing the moments of the size distribution,
such as the average size of (finite) connected components~$S(z)$. First let us work below the phase transition, such that there is no giant component,
and~$H_0(1) = H_1(1) = 1$. Using~\eqref{H0 implicit}, we obtain
\be
S(z) = H_0'(1) = 1 + G_0'(1) H_1'(1)\,.
\label{mean s}
\ee
On the other hand, from~\eqref{H1 implicit} we have~$H_1'(1) = 1 + G_1'(1) H_1'(1)$, which implies~$H_1'(1) = \frac{1}{1 - G_1'(1)}$. Thus we obtain
\be
S(z) =  1 + \frac{G_0'(1)}{1 - G_1'(1)} = 1 + \frac{z^2}{z - z_2}\,,
\label{mean s done}
\ee
where in the last step we have used~\eqref{z} and~\eqref{z2}. It is clear that the percolation threshold where a giant component emerges is given by the condition
\be
z_2 = z  \qquad (\text{percolation})\,.
\label{perc cond}
\ee
(Notice that this is identical to~\eqref{perc cond directed}, hence the percolation structure matches that of the directed case.) This result holds for any degree distribution~$p_k$ with finite mean and variance. More generally, above the phase transition~$S(z)$ gives the average size of finite clusters, {\it i.e.}, excluding the giant component.
Using~\eqref{P(z) def}, the generalization of~\eqref{mean s} is
\be
S(z) = \frac{H_0'(1)}{H_0(1)} = 1 + \frac{ z H_1^2(1) }{\big(1-P(z)\big)\Big[1- G_1'(H_1(1)) \Big]}\,,
\label{mean s gen}
\ee
where~$H_1(1)=u$ is the probability that there is no infinite cluster going down an edge. It is the smallest solution of~$u = G_1(u)$, with~$ 0\le u \le 1$. 

\begin{figure}[h]
\center 
\includegraphics[scale=0.25]{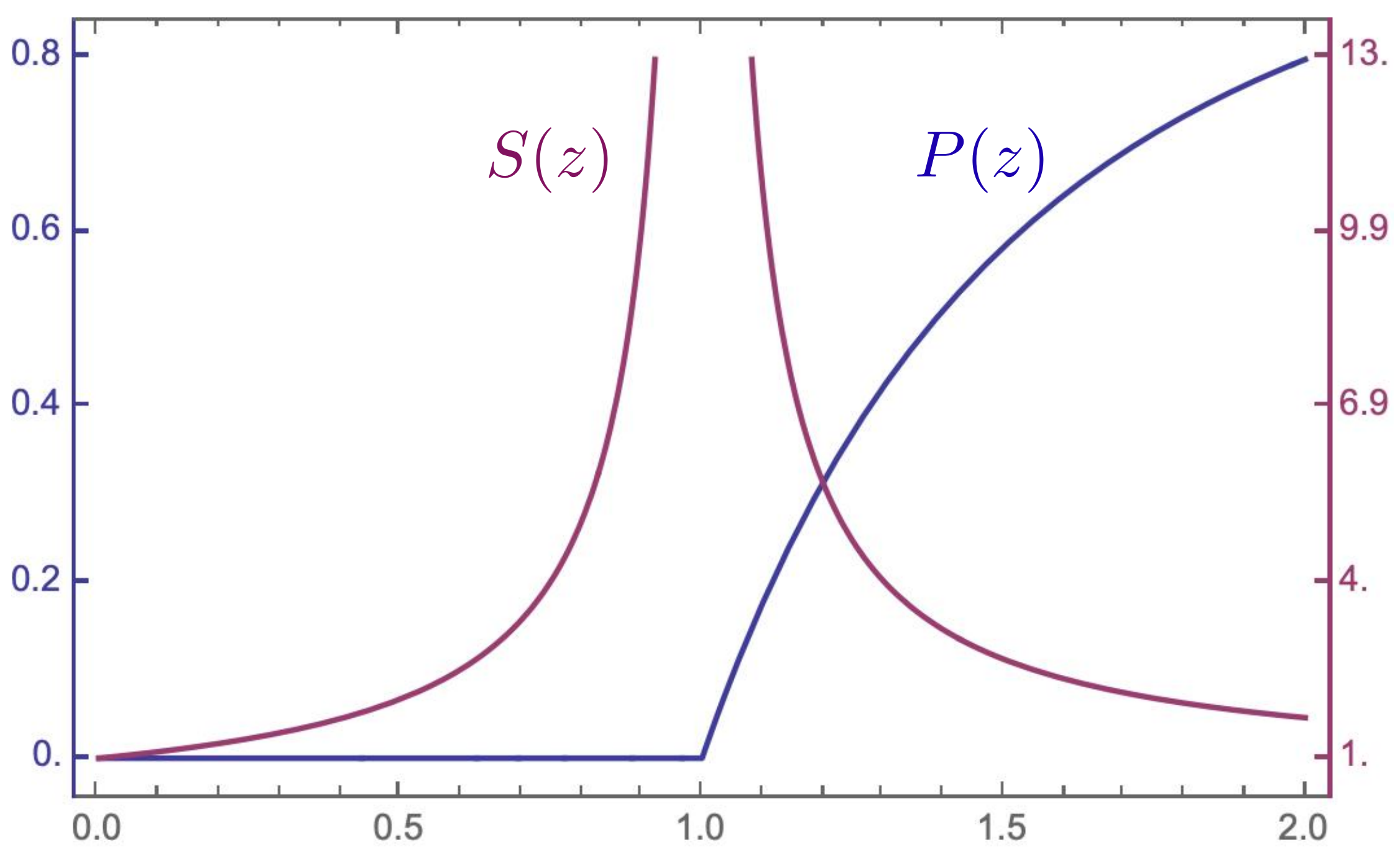}
\caption{Percolation criticality for Erd\"os-R\'enyi graphs. Blue curve: The fraction of vertices~$P(z)$ belonging to the giant component, given by~\eqref{P(z) ER}. Purple curve: The average size of finite clusters,~$S(z)$, excluding the giant component, given by~\eqref{eqn:clusmeansizeER}.}
\label{fig:ER}
\end{figure}

As an explicit example, consider Erd\"os-R\'enyi graphs. Combining~\eqref{z2z Poisson} and~\eqref{perc cond}, we recover the classic result that percolation
occurs at a critical mean degree~$z_{\rm c} = 1$, and corresponding critical probability~$p_{\rm c} \simeq \frac{1}{N}$.
Using~\eqref{G0G1ER}, the implicit equations~\eqref{H1 implicit} and~\eqref{H0 implicit} are readily solved in terms of the Lambert~${\cal W}$ function:
\be
H_0(x) = H_1(x) = - \frac{1}{z}{\cal W}\big( -z x {\rm e}^{-z}\big)  \qquad (\text{ER graphs})\,.
\ee
The fraction of vertices belonging to a giant cluster~\eqref{P(z) def} is given by 
\be
P(z) = 1- H_0(1)  =1 + \frac{{\cal W}(-z e^{-z})}{z}\,.
\label{P(z) ER}
\ee
Equivalently,~$P(z)$ satisfies the well-known implicit relation~$P(z) = 1- \exp\big(-z P(z)\big)$. As shown in Fig.~\ref{fig:ER} (blue curve),~$P(z)$ vanishes identically below percolation, starts growing at the percolation threshold~$z_{\rm c} = 1$, and approaches unity for large~$z$. Using the general expression~\eqref{mean s gen}, the average size of finite clusters in this case is given by
\begin{equation}
  S(z) = \frac{1}{ 1- z \big(1-P(z)\big)}\,.
\label{eqn:clusmeansizeER}
\end{equation} 
As shown in Fig.~\ref{fig:ER} (purple curve), this diverges at criticality.

\subsection{Tail component size distribution near percolation threshold}

Close to the percolation, the tail of the component size distribution~$P_s$ takes the form~\eqref{Ps tail}. Analogously to the directed case, 
at criticality this distribution has a power-law tail, with the same~$-3/2$ critical exponent~\cite{PhysRevE.64.026118}:
\be
P_s \sim s^{-3/2}\,.
\label{P_s crit}
\ee
See Appendix~\ref{critical exp} for a brief proof of this result. A related quantity is~$n_s$, the probability distribution for the number of clusters of size~$s$, which also exhibits a power-law tail~$n_s \sim s^{-5/2}$ at criticality. In general,~$n_s$ is related to~$P_s$ by~$P_s = s n_s$, where the factor of~$s$ accounts for the fact that one is~$s$ times more likely to randomly pick a node belonging to a cluster of size~$s$ than an isolated node. 
The giant component is of size~$N^{2/3}$ at criticality. All other components have size~${\cal O}(\log N)$. Above the percolation threshold, the giant component grows to encompass an~${\cal O}(1)$ fraction of all the nodes.

\section{Critical Exponent for Component Size Distribution} 
\label{critical exp}

In the main text we claimed that, for undirected random graphs, the distribution~$P_s$ of (finite) component sizes from a randomly-chosen vertex develops a power-law tail at criticality.
In this Appendix, for completeness we briefly outline the proof of this result, closely following~\cite{PhysRevE.64.026118}. We focus on the Erd\"os-R\'enyi universality class, which includes all degree
distributions with finite mean and variance.

Near the percolation threshold, the tail of~$P_s$ takes the form
\be
P_s \sim s^{-\tau} {\rm e}^{-s/s_{\rm max}}\,.
\label{Ps tail app}
\ee
As usual, the asymptotic behavior of~$P_s$ is encoded in the behavior of its generating function~$H_0(x)$ near its radius of convergence~$|x_\star|$~\cite{wilf}.
(The radius of convergence is defined as the singularity in~$H_0(x)$ closest to the origin.) Concretely, the correlation length~$s_{\rm max}$ is given by
\be
s_{\rm max} = \frac{1}{\log |x_\star|}\,.
\label{s max app}
\ee
Given the implicit relation~\eqref{H0 implicit}, and using the fact that the first singularity in~$G_0(x)$
lies outside the unit circle, one can deduce that~$x_\star$ also corresponds to the singularity in~$H_1(x)$ closest
to the origin. 

With these facts at hand, consider the implicit relation~\eqref{H1 implicit} for~$H_1$:
\be
H_1(x) = xG_1\big(H_1(x)\big)\,.
\label{H1 implicit app}
\ee
Letting~$w \equiv H_1(x)$, this can be expressed as
\be
x = H^{-1}_1(w) = \frac{w}{G_1(w)}\,.
\label{xw}
\ee
The singularity at~$x_\star$ entails that~$\frac{{\rm d}x}{{\rm d}H_1}\big\vert_{x = x_\star}= 0$, and therefore~$\frac{{\rm d}H_1^{-1}(w)}{{\rm d}w}\big\vert_{w=w_\star} = 0$. Using~\eqref{xw},
this last statement implies
\be
G_1(w_\star) = w_\star G_1'(w_\star)\,.
\label{G1 w*}
\ee
Given a solution~$w_\star$ to~\eqref{G1 w*}, we can solve~\eqref{xw} to find the corresponding~$x_\star$:
\be
x_\star = \frac{w_\star}{G_1(w_\star)}\,,
\label{x*}
\ee
and thus~$s_{\rm max}$ via~\eqref{s max app}. It should be stressed that~\eqref{G1 w*} need not have a solution. When it does not, then the asymptotic behavior of~$P_s$ is not of the form~\eqref{Ps tail app}.
We will see, however, that a solution to~\eqref{G1 w*} exists close to the phase transition.

Indeed, at the percolation threshold, defined by~$G_1'(1) = 1$,~\eqref{G1 w*} and~\eqref{x*} are solved by
\be
x_\star = w_\star = 1 \qquad \text{at percolation}\,.
\ee
Correspondingly,~$s_{\rm max} \rightarrow \infty$. To determine the power-law for~$P_s$ at criticality, let us expand~\eqref{xw} around~$x_\star = w_\star = 1$:
\be
x = H^{-1}_1(w)  = \frac{1 + (w - 1) }{G_1\big(1 + (w-1)\big)} \simeq 1 - \frac{1}{2} G''_1(1) \left(w - 1\right)^2 + \ldots 
\label{xw expand}
\ee
where we have used~$G_1(1) = G_1'(1) = 1$. The derivation clearly relies on~$G_1''(1)$ being finite. From~\eqref{G1}, this amounts to assuming that the first three moments of the degree distribution are finite.
This is the case, in particular, for the Poissonian distribution, but not for scale-free graphs. It follows from~\eqref{xw expand} that
\be
|1 - w| = \sqrt{\frac{2}{\left\vert G_1''(1)\right\vert}} \sqrt{1-x}\,. 
\ee
Therefore, since~$w = H_1(x)$, we deduce that the singular behavior of~$H_1$, and therefore that of~$H_0(x)$ as well, near~$x = 1$ is given by
\be
H_0(x) \sim H_1(x) \sim (1 - x)^\beta\,;\qquad ~~\text{with}~~\beta = \frac{1}{2}\,.
\label{H0beta}
\ee

The critical exponent~$\beta = 1/2$ is related to the critical exponent~$\tau$ for~$P_s$ as follows. On the one hand, from~\eqref{H0beta} we have
\be
\beta =  1 + \lim_{x\rightarrow 1} (x-1) \frac{H_0''(x)}{H'_0(x)} \,.
\ee
On the other hand, the asymptotic form~\eqref{Ps tail app} implies that~$H_0(x)$ can be expressed as
\be
H_0(x) = \sum_{s = 0}^{a-1} P_s x^s + C \sum_{s = a}^\infty s^{-\tau} {\rm e}^{-s/s_{\rm max}}x^s + \ldots
\ee
where~$C$ is a constant, and~$a$ is sufficiently large. The singular behavior as~$x\rightarrow 1$ is encoded in the infinite sum. Focusing on this term we have
\be
\beta \simeq  1  +  \lim_{x\rightarrow 1}  \frac{x-1}{x}  \frac{\sum\limits_{s = a}^\infty s^{2-\tau} x^{s-1}}{\sum\limits_{s = a}^\infty s^{1-\tau} x^{s-1}}\,.
\label{beta lim}
\ee
In the limit of large~$a$, {\it i.e.}, large~$s$, the infinite sums can be approximated as integrals:
\bea
\nonumber
\frac{\sum\limits_{s = a}^\infty s^{2-\tau} x^{s-1}}{\sum\limits_{s = a}^\infty s^{1-\tau} x^{s-1}} &\simeq & \frac{\int\limits_a^\infty {\rm d}s\, s^{2-\tau}x^s}{\int\limits_a^\infty {\rm d}s\, s^{1-\tau}x^s} \\
\nonumber
& = & - \frac{1}{\log x} \, \frac{\int\limits_{-a\log x}^\infty {\rm d} t \, t^{2-\tau} {\rm e}^{-t}}{\int\limits_{-a\log x}^\infty {\rm d} t \, t^{1-\tau} {\rm e}^{-t}}  \\
&=&  - \frac{1}{\log x}\, \frac{\Gamma(3-\tau,-a \log x)}{\Gamma(2 - \tau, -a \log x)} \,.
\eea
Substituting into~\eqref{beta lim} and taking the limit~$x \rightarrow 1$, we obtain
\be
\beta = \tau - 1 \,.
\ee
Since~$\beta = 1/2$, it follows that~$\tau =  3/2$, which proves~\eqref{alpha def}.

\section{Bethe Lattice and its Percolation Structure}
\label{Bethe lattice sec}

As a second approach to percolation on the landscape, the underlying graph topology is modeled as a Bethe lattice (or Cayley tree). This approach is appropriate whenever transitions are strictly ``local" in field space, {\it i.e.},
they are non-negligible only between ``nearest-neighbor" vacua, while transitions to more distant vacua can be safely ignored. 

\begin{figure}[h]
\center 
\includegraphics[scale=0.7]{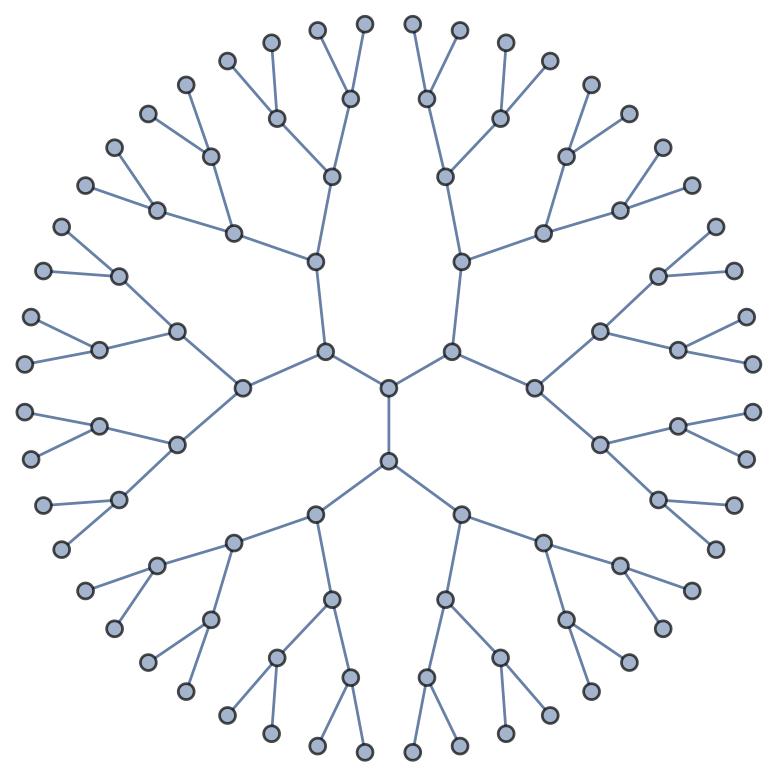}
\caption{A Bethe lattice with degree $k=3$.}
\label{fig:bethe}
\end{figure}

\subsection{Undirected Bethe lattice} 

A Bethe lattice is loop-free graph where all vertices have the same degree~$k$. Thus it is a tree with branching number~$k-1$. An example is given in Fig.~\ref{fig:bethe}. In the undirected case, there is nothing special about the central vertex, since every vertex can be viewed as the root. 

Thanks to the self-similar structure of this graph, the percolation problem is easily solved. As usual, one assigns a probability~$p$ for each edge to be included. The percolation phase transition, defined by the emergence of a giant cluster, occurs at a critical~$p_{\rm c}$. More precisely, in the~$N\rightarrow\infty$ limit, the probability~$P(p)$ that a vertex is part of an infinite cluster ({\it i.e.}, connected to infinitely-many vertices) vanishes for~$p < p_{\rm c}$ and is finite for~$p > p_{\rm c}$.  Clearly~$P(p)$ satisfies
\begin{equation}
P(p) =  1- Q^k(p)\,,
\label{Bethe P reln}
\end{equation}
where~$Q(p)$ is the probability that a neighboring vertex does not connect to infinity on the other side. Due to the similarity of the next layer of neighbor,~$Q(p)$ satisfies the implicit relation,
\begin{equation}
Q = \sum_{\ell=0}^{k-1} {k-1 \choose \ell} (1-p)^{k-1-\ell} (pQ)^\ell = \Big(1-p (1- Q)\Big)^{k-1}\,,
\label{Bethe Q reln}
\end{equation} 
where the first equality accounts for all combinations of connecting to the next layer. 

The critical probability~$p_{\rm c}$ of percolation is defined such that~$1>Q(p>p_{\rm c})>0$, {\it i.e.}, the boundary of~$p$ in which~$Q(p)$ is non-zero. 
To obtain an expression for~$p_{\rm c}$, let us work near the percolation threshold, where~$1-Q\ll 1$. Expanding the implicit relation~\eqref{Bethe Q reln} for~$1-Q\ll 1$, we obtain
\begin{equation}
 1-Q \simeq (k-1)p(1-Q) \quad  \Longrightarrow \quad p_{\rm c} = \frac{1}{k-1}\,.
\label{pc Bethe}
\end{equation}
Thus at percolation criticality each vertex has one connection to the next layer on average. (For~$p < p_{\rm c}$, the only solution to~\eqref{pc Bethe}, and more generally~\eqref{Bethe Q reln}, is~$Q(p) = 1$, corresponding to~$P(p) = 0$.) 

The next quantity of interest is the average cluster size,~$S(p)$, for $p\leq p_{\rm c}$. It satisfies 
\begin{equation}
 S(p) = 1 + k T(p)\,,
\label{ST Bethe}
\end{equation}
where~$T(p)$ is the expected size of the sub-branch going down an edge.
Again, due to the self-similar structure, we have the implicit relation
\begin{equation}
 T = p\Big(1+ (k-1) T\Big) \,.
\end{equation}
The factor~$p$ accounts for the probability of connection, and~$1 + (k-1) T$ is the vertex itself plus the expected size of the~$k-1$ vertices at the next layer.
Solving for~$T$ and substituting~\eqref{pc Bethe} gives
\be
T(p) = \frac{p_{\rm c} p}{p_{\rm c} -p}\,.
\ee
Substituting into~\eqref{ST Bethe} gives  
\begin{equation} \label{eqn:S(p)}
 S(p) = \frac{p_{\rm c} (1+p)}{p_{\rm c} - p}\qquad (p \leq p_{\rm c})\,.
\end{equation}
As expected, the average cluster size diverges as~$p\rightarrow p_{\rm c}$ from below. The probability~$P(p)$ and the average size of cluster~$S(p)$ are plotted as a function of~$p$ in Fig.~\ref{fig:bethe percolation}. 

\begin{figure}[h]
\center 
\includegraphics[scale=0.6]{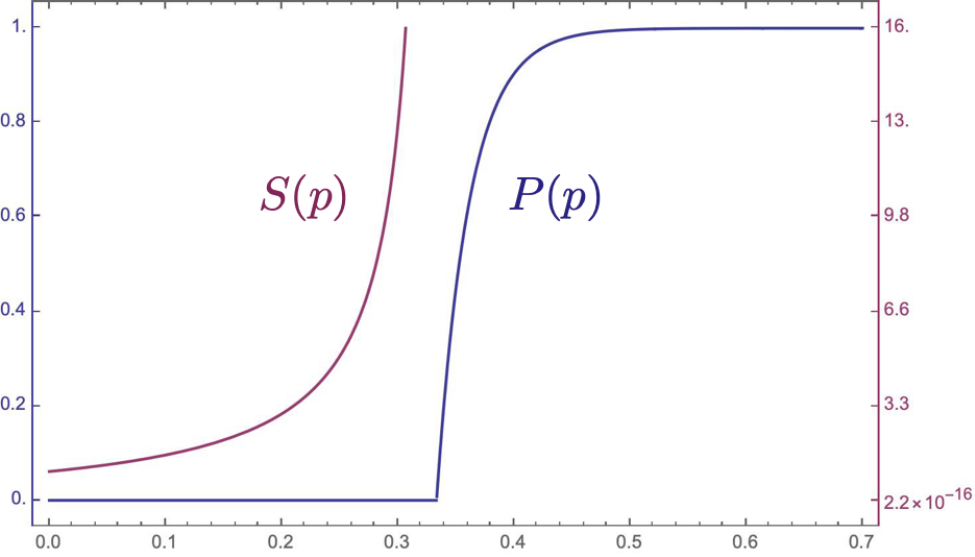}
\caption{The probability~$P(p)$ for a vertex to belong to an infinite cluster (blue) and the average cluster size~$S(p)$ (orange) as a function of~$p$ for an undirected Bethe lattice with degree~$k = 4$, corresponding to~$p_{\rm c} = 1/3$. }
\label{fig:bethe percolation}
\end{figure}

%

Near the percolation threshold, the tail of the component size distribution~$P_s$ takes the same form as~\eqref{Ps tail app}. Remarkably, it can be shown that it has the same critical exponent as
Erd\"os-R\'enyi graphs,
\be
P_s \sim s^{-3/2} \,,
\label{-3/2 Bethe}
\ee
for any~$k$ ~\cite{percolationtextbook,Essamreview}. To show this, one starts with the quantity 
\begin{align}
   n_s(p): \mbox{the number of $s$-size cluster per site}\,,
\end{align}
such that $sn_s(p)$ is the fraction of sites belonging to an $s$-size cluster. In general, the chance of forming an $s$-size cluster is 
\begin{align}
  n(s,p) = g_{s,\alpha} p^{s-1}(1-p)^{\alpha}\,,
\end{align}
where $p^{s-1}$ accounts for $s-1$ connected edges, $(1-p)^{\alpha}$ accounts for $\alpha$ disconnected edges at the ``boundary" of the cluster, and  $g_{s,\alpha}$ is a combinatoric factor. One can easily figure out that $\alpha =s(k-1) +2$, therefore 
\begin{align}
 \frac{n(s,p)}{n(s,p_c)} & = \frac{p_c}{p} \left( \frac{1-p}{1-p_c}\right)^2 \exp \left[s \ln\left( (k-1)p \left(\frac{(k-1)(1-p)}{k-2}\right)^{k-2} \right) \right] \nonumber \\
  & \sim \frac{p_c}{p} \left( \frac{1-p}{1-p_c}\right)^2 \exp \left( -\frac{s}{\ell(p)} \right) \,,
\end{align}
where we only focus on the proximity of the percolation threshold, and 
\begin{align}
 \frac{1}{\ell(p)} = \frac{1}{2p_c^2(1-p_c)} (p-p_c)^2 + {\cal O}\left( (p-p_c)^3 \right)\,.
\label{l(p)}
\end{align}
Furthermore we can approximate that, for large $s$ and $k>2$, 
\begin{align}
   n(s,p) \propto s^{-\sigma} \exp \left( -\frac{s}{\ell(p)}\right)\,.
\end{align}

Therefore the average cluster size $S(p)$, can be approximated as
\begin{align}
 S(p) = \sum_s s^2 n(s,p) & \approx \sum_s s^{2-\sigma}\exp \left( -\frac{s}{\ell(p)}\right)  \nonumber \\
 & \simeq \int^{\infty}_1\rd s \, s^{2-\sigma}\exp \left( -\frac{s}{\ell(p)}\right) \nonumber\\
 & \simeq \big(\ell(p)\big)^{3-\sigma} \Gamma(3-\tau) \nonumber\\
 & \propto (p-p_c)^{2\sigma -6}\,.
\end{align} 
At the same time we know that $S(p) \propto (p_c-p)^{-1} $ from \eqref{eqn:S(p)}, therefore 
\begin{align}
 \sigma = \frac{5}{2}\,. 
\end{align}
By definition, the chance of a randomly chosen site belonging to an $s$-size cluster is then, 
\begin{align}
  P_s(p) = s n(s,p)  \propto s^{-3/2}\exp \left( -\frac{s}{\ell(p)} \right)\,. 
\end{align}
At criticality,~$\ell(p)$ diverges per~\eqref{l(p)}, and we recover~\eqref{-3/2 Bethe}.

\subsection{Directed Bethe lattice} 

In the situation of interest, the downward approximation implies that the graph is a {\it directed Bethe lattice}. This reflects the fact that low-lying (high-lying) vacua have mostly incoming (outgoing) edges. Choosing a fixed vertex as the center, we denote by~$r$ the probability that an edge is directed outward from this center, and by~$s = 1-  r$ the probability of the edge being directed inward. Thus the probability of including an inward edge is~$p r$. For general~$r$, vertices are no longer indistinguishable since the chosen center vertex is indeed the center. (The exception is~$r =\frac{1}{2}$, where every vertex is still statistically the same.) Nevertheless, away from the center vertex, layers after layers still have the same self-similarity structure.

Certain properties are insensitive to~$r$, such as the emergence of the giant component. One can ask about the probability for the existence of a directed path from the center vertex to infinity. The derivation of~$P(p)$ in the undirected case applies just as well to the directed case with the replacement~$p \to pr$. Therefore the probability for the existence of a path {\it to} infinity and the probability for a path {\it from}  infinity are respectively given by
\bea
\nonumber
\mbox{a path to }\infty &: & \quad P(pr)\,; \\
\mbox{a path from }\infty &:& \quad P(ps)\,.
\eea
Similarly, the average size of the outward- and inward-pointing clusters are respectively given by~$S(pr)$ and~$S(ps)$.
Obviously their sum must be strictly smaller than the average cluster size in the undirected case, 
\be
S(pr) +S(ps) - 1 < S(p)\,,
\ee
below the percolation threshold,~$p\leq p_{\rm c}$.

\end{appendices}

\bibliographystyle{utphys}
\bibliography{directed_percolation_v5.bib}

\providecommand{\href}[2]{#2}\begingroup\raggedright\begin{thebibliography}{100}

\bibitem{Degrassi:2012ry}
G.~Degrassi, S.~Di~Vita, J.~Elias-Miro, J.~R. Espinosa, G.~F. Giudice,
  G.~Isidori, and A.~Strumia, ``{Higgs mass and vacuum stability in the
  Standard Model at NNLO},''
  \href{http://dx.doi.org/10.1007/JHEP08(2012)098}{{\em JHEP} {\bfseries 08}
  (2012) 098}, \href{http://arxiv.org/abs/1205.6497}{{\ttfamily arXiv:1205.6497
  [hep-ph]}}.

\bibitem{Buttazzo:2013uya}
D.~Buttazzo, G.~Degrassi, P.~P. Giardino, G.~F. Giudice, F.~Sala, A.~Salvio,
  and A.~Strumia, ``{Investigating the near-criticality of the Higgs boson},''
  \href{http://dx.doi.org/10.1007/JHEP12(2013)089}{{\em JHEP} {\bfseries 12}
  (2013) 089}, \href{http://arxiv.org/abs/1307.3536}{{\ttfamily arXiv:1307.3536
  [hep-ph]}}.

\bibitem{Lalak:2014qua}
Z.~Lalak, M.~Lewicki, and P.~Olszewski, ``{Higher-order scalar interactions and
  SM vacuum stability},'' \href{http://dx.doi.org/10.1007/JHEP05(2014)119}{{\em
  JHEP} {\bfseries 05} (2014) 119},
  \href{http://arxiv.org/abs/1402.3826}{{\ttfamily arXiv:1402.3826 [hep-ph]}}.

\bibitem{Andreassen:2014gha}
A.~Andreassen, W.~Frost, and M.~D. Schwartz, ``{Consistent Use of the Standard
  Model Effective Potential},''
  \href{http://dx.doi.org/10.1103/PhysRevLett.113.241801}{{\em Phys. Rev.
  Lett.} {\bfseries 113} no.~24, (2014) 241801},
  \href{http://arxiv.org/abs/1408.0292}{{\ttfamily arXiv:1408.0292 [hep-ph]}}.

\bibitem{Branchina:2014rva}
V.~Branchina, E.~Messina, and M.~Sher, ``{Lifetime of the electroweak vacuum
  and sensitivity to Planck scale physics},''
  \href{http://dx.doi.org/10.1103/PhysRevD.91.013003}{{\em Phys. Rev. D}
  {\bfseries 91} (2015) 013003},
  \href{http://arxiv.org/abs/1408.5302}{{\ttfamily arXiv:1408.5302 [hep-ph]}}.

\bibitem{Bednyakov:2015sca}
A.~Bednyakov, B.~Kniehl, A.~Pikelner, and O.~Veretin, ``{Stability of the
  Electroweak Vacuum: Gauge Independence and Advanced Precision},''
  \href{http://dx.doi.org/10.1103/PhysRevLett.115.201802}{{\em Phys. Rev.
  Lett.} {\bfseries 115} no.~20, (2015) 201802},
  \href{http://arxiv.org/abs/1507.08833}{{\ttfamily arXiv:1507.08833
  [hep-ph]}}.

\bibitem{Iacobellis:2016eof}
G.~Iacobellis and I.~Masina, ``{Stationary configurations of the Standard Model
  Higgs potential: electroweak stability and rising inflection point},''
  \href{http://dx.doi.org/10.1103/PhysRevD.94.073005}{{\em Phys. Rev. D}
  {\bfseries 94} no.~7, (2016) 073005},
  \href{http://arxiv.org/abs/1604.06046}{{\ttfamily arXiv:1604.06046
  [hep-ph]}}.

\bibitem{Andreassen:2017rzq}
A.~Andreassen, W.~Frost, and M.~D. Schwartz, ``{Scale Invariant Instantons and
  the Complete Lifetime of the Standard Model},''
  \href{http://dx.doi.org/10.1103/PhysRevD.97.056006}{{\em Phys. Rev. D}
  {\bfseries 97} no.~5, (2018) 056006},
  \href{http://arxiv.org/abs/1707.08124}{{\ttfamily arXiv:1707.08124
  [hep-ph]}}.

\bibitem{Khoury:2021zao}
J.~Khoury and T.~Steingasser, ``{Gauge hierarchy from electroweak vacuum
  metastability},'' \href{http://dx.doi.org/10.1103/PhysRevD.105.055031}{{\em
  Phys. Rev. D} {\bfseries 105} no.~5, (2022) 055031},
  \href{http://arxiv.org/abs/2108.09315}{{\ttfamily arXiv:2108.09315
  [hep-ph]}}.

\bibitem{Steingasser:2023ugv}
T.~Steingasser and D.~I. Kaiser, ``{Higgs Criticality beyond the Standard
  Model},'' \href{http://arxiv.org/abs/2307.10361}{{\ttfamily arXiv:2307.10361
  [hep-ph]}}.

\bibitem{Giudice:2006sn}
G.~Giudice and R.~Rattazzi, ``{Living Dangerously with Low-Energy
  Supersymmetry},''
  \href{http://dx.doi.org/10.1016/j.nuclphysb.2006.07.031}{{\em Nucl. Phys. B}
  {\bfseries 757} (2006) 19--46},
  \href{http://arxiv.org/abs/hep-ph/0606105}{{\ttfamily arXiv:hep-ph/0606105}}.

\bibitem{Friedrich:1987}
H.~Friedrich, ``{On the existence of n-geodesically complete or future complete
  solutions of Einstein's field equations with smooth asymptotic structure},''
  {\em Communications in Mathematical Physics} {\bfseries 107} no.~4, (Dec.,
  1986) 587--609.

\bibitem{Bizon:2011gg}
P.~Bizon and A.~Rostworowski, ``{On weakly turbulent instability of anti-de
  Sitter space},'' \href{http://dx.doi.org/10.1103/PhysRevLett.107.031102}{{\em
  Phys. Rev. Lett.} {\bfseries 107} (2011) 031102},
  \href{http://arxiv.org/abs/1104.3702}{{\ttfamily arXiv:1104.3702 [gr-qc]}}.

\bibitem{Steinhardt:1982kg}
P.~J. Steinhardt, ``{Natural Inflation},'' in {\em {Nuffield Workshop on the
  Very Early Universe}}, pp.~251--266.
\newblock 7, 1982.

\bibitem{Vilenkin:1983xq}
A.~Vilenkin, ``{The Birth of Inflationary Universes},''
  \href{http://dx.doi.org/10.1103/PhysRevD.27.2848}{{\em Phys. Rev. D}
  {\bfseries 27} (1983) 2848}.

\bibitem{Linde:1986fc}
A.~D. Linde, ``{Eternal Chaotic Inflation},''
  \href{http://dx.doi.org/10.1142/S0217732386000129}{{\em Mod. Phys. Lett. A}
  {\bfseries 1} (1986) 81}.

\bibitem{Linde:1986fd}
A.~D. Linde, ``{Eternally Existing Selfreproducing Chaotic Inflationary
  Universe},'' \href{http://dx.doi.org/10.1016/0370-2693(86)90611-8}{{\em Phys.
  Lett. B} {\bfseries 175} (1986) 395--400}.

\bibitem{Starobinsky:1986fx}
A.~A. Starobinsky, ``{Stochastic de Sitter (Inflationary) Stage in the Early
  Universe},'' \href{http://dx.doi.org/10.1007/3-540-16452-9_6}{{\em Lect.
  Notes Phys.} {\bfseries 246} (1986) 107--126}.

\bibitem{Bousso:2000xa}
R.~Bousso and J.~Polchinski, ``{Quantization of four form fluxes and dynamical
  neutralization of the cosmological constant},''
  \href{http://dx.doi.org/10.1088/1126-6708/2000/06/006}{{\em JHEP} {\bfseries
  06} (2000) 006}, \href{http://arxiv.org/abs/hep-th/0004134}{{\ttfamily
  arXiv:hep-th/0004134}}.

\bibitem{Kachru:2003aw}
S.~Kachru, R.~Kallosh, A.~D. Linde, and S.~P. Trivedi, ``{De Sitter vacua in
  string theory},'' \href{http://dx.doi.org/10.1103/PhysRevD.68.046005}{{\em
  Phys. Rev. D} {\bfseries 68} (2003) 046005},
  \href{http://arxiv.org/abs/hep-th/0301240}{{\ttfamily arXiv:hep-th/0301240}}.

\bibitem{Ashok:2003gk}
S.~Ashok and M.~R. Douglas, ``{Counting flux vacua},''
  \href{http://dx.doi.org/10.1088/1126-6708/2004/01/060}{{\em JHEP} {\bfseries
  01} (2004) 060}, \href{http://arxiv.org/abs/hep-th/0307049}{{\ttfamily
  arXiv:hep-th/0307049}}.

\bibitem{Linde:1993nz}
A.~D. Linde and A.~Mezhlumian, ``{Stationary universe},''
  \href{http://dx.doi.org/10.1016/0370-2693(93)90187-M}{{\em Phys. Lett. B}
  {\bfseries 307} (1993) 25--33},
  \href{http://arxiv.org/abs/gr-qc/9304015}{{\ttfamily arXiv:gr-qc/9304015}}.

\bibitem{Linde:1993xx}
A.~D. Linde, D.~A. Linde, and A.~Mezhlumian, ``{From the Big Bang theory to the
  theory of a stationary universe},''
  \href{http://dx.doi.org/10.1103/PhysRevD.49.1783}{{\em Phys. Rev. D}
  {\bfseries 49} (1994) 1783--1826},
  \href{http://arxiv.org/abs/gr-qc/9306035}{{\ttfamily arXiv:gr-qc/9306035}}.

\bibitem{Guth:2007ng}
A.~H. Guth, ``{Eternal inflation and its implications},''
  \href{http://dx.doi.org/10.1088/1751-8113/40/25/S25}{{\em J. Phys. A}
  {\bfseries 40} (2007) 6811--6826},
  \href{http://arxiv.org/abs/hep-th/0702178}{{\ttfamily arXiv:hep-th/0702178}}.

\bibitem{Freivogel:2011eg}
B.~Freivogel, ``{Making predictions in the multiverse},''
  \href{http://dx.doi.org/10.1088/0264-9381/28/20/204007}{{\em Class. Quant.
  Grav.} {\bfseries 28} (2011) 204007},
  \href{http://arxiv.org/abs/1105.0244}{{\ttfamily arXiv:1105.0244 [hep-th]}}.

\bibitem{Khoury:2022ish}
J.~Khoury and S.~S.~C. Wong, ``{Bayesian reasoning in eternal inflation: A
  solution to the measure problem},''
  \href{http://dx.doi.org/10.1103/PhysRevD.108.023506}{{\em Phys. Rev. D}
  {\bfseries 108} no.~2, (2023) 023506},
  \href{http://arxiv.org/abs/2205.11524}{{\ttfamily arXiv:2205.11524
  [hep-th]}}.

\bibitem{Borde:2001nh}
A.~Borde, A.~H. Guth, and A.~Vilenkin, ``{Inflationary space-times are
  incompletein past directions},''
  \href{http://dx.doi.org/10.1103/PhysRevLett.90.151301}{{\em Phys. Rev. Lett.}
  {\bfseries 90} (2003) 151301},
  \href{http://arxiv.org/abs/gr-qc/0110012}{{\ttfamily arXiv:gr-qc/0110012}}.

\bibitem{Garriga:2005av}
J.~Garriga, D.~Schwartz-Perlov, A.~Vilenkin, and S.~Winitzki, ``{Probabilities
  in the inflationary multiverse},''
  \href{http://dx.doi.org/10.1088/1475-7516/2006/01/017}{{\em JCAP} {\bfseries
  01} (2006) 017}, \href{http://arxiv.org/abs/hep-th/0509184}{{\ttfamily
  arXiv:hep-th/0509184}}.

\bibitem{GarciaBellido:1993wn}
J.~Garcia-Bellido, A.~D. Linde, and D.~A. Linde, ``{Fluctuations of the
  gravitational constant in the inflationary Brans-Dicke cosmology},''
  \href{http://dx.doi.org/10.1103/PhysRevD.50.730}{{\em Phys. Rev. D}
  {\bfseries 50} (1994) 730--750},
  \href{http://arxiv.org/abs/astro-ph/9312039}{{\ttfamily
  arXiv:astro-ph/9312039}}.

\bibitem{Vilenkin:1994ua}
A.~Vilenkin, ``{Predictions from quantum cosmology},''
  \href{http://dx.doi.org/10.1103/PhysRevLett.74.846}{{\em Phys. Rev. Lett.}
  {\bfseries 74} (1995) 846--849},
  \href{http://arxiv.org/abs/gr-qc/9406010}{{\ttfamily arXiv:gr-qc/9406010}}.

\bibitem{Garriga:1997ef}
J.~Garriga and A.~Vilenkin, ``{Recycling universe},''
  \href{http://dx.doi.org/10.1103/PhysRevD.57.2230}{{\em Phys. Rev. D}
  {\bfseries 57} (1998) 2230--2244},
  \href{http://arxiv.org/abs/astro-ph/9707292}{{\ttfamily
  arXiv:astro-ph/9707292}}.

\bibitem{Garriga:2001ri}
J.~Garriga and A.~Vilenkin, ``{A Prescription for probabilities in eternal
  inflation},'' \href{http://dx.doi.org/10.1103/PhysRevD.64.023507}{{\em Phys.
  Rev. D} {\bfseries 64} (2001) 023507},
  \href{http://arxiv.org/abs/gr-qc/0102090}{{\ttfamily arXiv:gr-qc/0102090}}.

\bibitem{Bousso:2006ev}
R.~Bousso, ``{Holographic probabilities in eternal inflation},''
  \href{http://dx.doi.org/10.1103/PhysRevLett.97.191302}{{\em Phys. Rev. Lett.}
  {\bfseries 97} (2006) 191302},
  \href{http://arxiv.org/abs/hep-th/0605263}{{\ttfamily arXiv:hep-th/0605263}}.

\bibitem{Friedrich:2022tqk}
B.~Friedrich, A.~Hebecker, M.~Salmhofer, J.~C. Strauss, and J.~Walcher, ``{A
  local Wheeler-DeWitt measure for the string landscape},''
  \href{http://dx.doi.org/10.1016/j.nuclphysb.2023.116230}{{\em Nucl. Phys. B}
  {\bfseries 992} (2023) 116230},
  \href{http://arxiv.org/abs/2205.09772}{{\ttfamily arXiv:2205.09772
  [hep-th]}}.

\bibitem{Denef:2017cxt}
F.~Denef, M.~R. Douglas, B.~Greene, and C.~Zukowski, ``{Computational
  complexity of the landscape II\textemdash{}Cosmological considerations},''
  \href{http://dx.doi.org/10.1016/j.aop.2018.03.013}{{\em Annals Phys.}
  {\bfseries 392} (2018) 93--127},
  \href{http://arxiv.org/abs/1706.06430}{{\ttfamily arXiv:1706.06430
  [hep-th]}}.

\bibitem{Khoury:2019yoo}
J.~Khoury and O.~Parrikar, ``{Search Optimization, Funnel Topography, and
  Dynamical Criticality on the String Landscape},''
  \href{http://dx.doi.org/10.1088/1475-7516/2019/12/014}{{\em JCAP} {\bfseries
  12} (2019) 014}, \href{http://arxiv.org/abs/1907.07693}{{\ttfamily
  arXiv:1907.07693 [hep-th]}}.

\bibitem{Khoury:2019ajl}
J.~Khoury, ``{Accessibility Measure for Eternal Inflation: Dynamical
  Criticality and Higgs Metastability},''
  \href{http://dx.doi.org/10.1088/1475-7516/2021/06/009}{{\em JCAP} {\bfseries
  06} (2021) 009}, \href{http://arxiv.org/abs/1912.06706}{{\ttfamily
  arXiv:1912.06706 [hep-th]}}.

\bibitem{Kartvelishvili:2020thd}
G.~Kartvelishvili, J.~Khoury, and A.~Sharma, ``{The Self-Organized Critical
  Multiverse},'' \href{http://dx.doi.org/10.1088/1475-7516/2021/02/028}{{\em
  JCAP} {\bfseries 02} (2021) 028},
  \href{http://arxiv.org/abs/2003.12594}{{\ttfamily arXiv:2003.12594
  [hep-th]}}.

\bibitem{Khoury:2021grg}
J.~Khoury and S.~S.~C. Wong, ``{Early-time measure in eternal inflation},''
  \href{http://dx.doi.org/10.1088/1475-7516/2022/05/031}{{\em JCAP} {\bfseries
  05} no.~05, (2022) 031}, \href{http://arxiv.org/abs/2106.12590}{{\ttfamily
  arXiv:2106.12590 [hep-th]}}.

\bibitem{Coleman:1977py}
S.~R. Coleman, ``{The Fate of the False Vacuum. 1. Semiclassical Theory},''
  \href{http://dx.doi.org/10.1103/PhysRevD.16.1248}{{\em Phys. Rev. D}
  {\bfseries 15} (1977) 2929--2936}. [Erratum: Phys.Rev.D 16, 1248 (1977)].

\bibitem{Callan:1977pt}
J.~Callan, Curtis~G. and S.~R. Coleman, ``{The Fate of the False Vacuum. 2.
  First Quantum Corrections},''
  \href{http://dx.doi.org/10.1103/PhysRevD.16.1762}{{\em Phys. Rev. D}
  {\bfseries 16} (1977) 1762--1768}.

\bibitem{Coleman:1980aw}
S.~R. Coleman and F.~De~Luccia, ``{Gravitational Effects on and of Vacuum
  Decay},'' \href{http://dx.doi.org/10.1103/PhysRevD.21.3305}{{\em Phys. Rev.
  D} {\bfseries 21} (1980) 3305}.

\bibitem{SchwartzPerlov:2006hi}
D.~Schwartz-Perlov and A.~Vilenkin, ``{Probabilities in the Bousso-Polchinski
  multiverse},'' \href{http://dx.doi.org/10.1088/1475-7516/2006/06/010}{{\em
  JCAP} {\bfseries 06} (2006) 010},
  \href{http://arxiv.org/abs/hep-th/0601162}{{\ttfamily arXiv:hep-th/0601162}}.

\bibitem{Olum:2007yk}
K.~D. Olum and D.~Schwartz-Perlov, ``{Anthropic prediction in a large toy
  landscape},'' \href{http://dx.doi.org/10.1088/1475-7516/2007/10/010}{{\em
  JCAP} {\bfseries 10} (2007) 010},
  \href{http://arxiv.org/abs/0705.2562}{{\ttfamily arXiv:0705.2562 [hep-th]}}.
  [Erratum: JCAP 10, E02 (2019)].

\bibitem{proteins1}
J.~D. {Bryngelson}, J.~N. {Onuchic}, N.~D. {Socci}, and P.~G. {Wolynes},
  ``{Funnels, Pathways and the Energy Landscape of Protein Folding: A
  Synthesis},'' {\em Proteins-Struct. Func. and Genetics} {\bfseries 21} (1995)
  167, \href{http://arxiv.org/abs/chem-ph/9411008}{{\ttfamily
  arXiv:chem-ph/9411008}}.

\bibitem{Doye2002}
J.~P.~K. Doye, ``{Network Topology of a Potential Energy Landscape: A Static
  Scale-Free Network},'' {\em Phys. Rev. Lett.} {\bfseries 88} (2002) 238701,
  \href{http://arxiv.org/abs/cond-mat/0201430}{{\ttfamily
  arXiv:cond-mat/0201430}}.

\bibitem{Doye2004}
J.~P.~K. Doye and C.~P. Massen, ``{Characterizing the network topology of the
  energy landscapes of atomic clusters},'' {\em J. Chem. Phys.} {\bfseries 122}
  (2005) 084105, \href{http://arxiv.org/abs/cond-mat/0411144}{{\ttfamily
  arXiv:cond-mat/0411144}}.

\bibitem{Doye2005}
J.~P.~K. Doye and C.~P. Massen, ``{Power-law distributions for the areas of the
  basins of attraction on a potential energy landscape},'' {\em Phys. Rev. E}
  {\bfseries 75} (2007) 037101,
  \href{http://arxiv.org/abs/cond-mat/0509185}{{\ttfamily
  arXiv:cond-mat/0509185}}.

\bibitem{DDNlossfunnel}
H.~Li, Z.~Xu, G.~Taylor, C.~Studer, and T.~Goldstein, ``{Visualizing the Loss
  Landscape of Neural Nets},''
  \href{http://arxiv.org/abs/1712.09913}{{\ttfamily arXiv:1712.09913 [cs.LG]}}.

\bibitem{TSP}
K.~D. Boese, A.~B. Boese, and S.~Muddu, ``{A new adaptive multi-start technique
  for combinatorial global optimizations},'' {\em Oper. Res. Lett.} {\bfseries
  16} (1994) 101--113.

\bibitem{Erdos:1959:pmd}
P.~Erd\"os and A.~R\'enyi, ``On random graphs i,'' {\em Publicationes
  Mathematicae Debrecen} {\bfseries 6} (1959) 290--297.

\bibitem{PhysRevE.64.026118}
M.~E.~J. Newman, S.~H. Strogatz, and D.~J. Watts, ``Random graphs with
  arbitrary degree distributions and their applications,''
  \href{http://dx.doi.org/10.1103/PhysRevE.64.026118}{{\em Phys. Rev. E}
  {\bfseries 64} (Jul, 2001) 026118}.

\bibitem{Bachlechner:2015gwa}
T.~C. Bachlechner, ``{Axionic Band Structure of the Cosmological Constant},''
  \href{http://dx.doi.org/10.1103/PhysRevD.93.023522}{{\em Phys. Rev. D}
  {\bfseries 93} no.~2, (2016) 023522},
  \href{http://arxiv.org/abs/1510.06388}{{\ttfamily arXiv:1510.06388
  [hep-th]}}.

\bibitem{Bachlechner:2017zpb}
T.~C. Bachlechner, K.~Eckerle, O.~Janssen, and M.~Kleban, ``{Multiple-axion
  framework},'' \href{http://dx.doi.org/10.1103/PhysRevD.98.061301}{{\em Phys.
  Rev. D} {\bfseries 98} no.~6, (2018) 061301},
  \href{http://arxiv.org/abs/1703.00453}{{\ttfamily arXiv:1703.00453
  [hep-th]}}.

\bibitem{Bachlechner:2018gew}
T.~C. Bachlechner, K.~Eckerle, O.~Janssen, and M.~Kleban, ``{Axion Landscape
  Cosmology},'' \href{http://dx.doi.org/10.1088/1475-7516/2019/09/062}{{\em
  JCAP} {\bfseries 09} (2019) 062},
  \href{http://arxiv.org/abs/1810.02822}{{\ttfamily arXiv:1810.02822
  [hep-th]}}.

\bibitem{Cespedes:2023jdk}
S.~Cespedes, S.~de~Alwis, F.~Muia, and F.~Quevedo, ``{Quantum Transitions,
  Detailed Balance, Black Holes and Nothingness},''
  \href{http://arxiv.org/abs/2307.13614}{{\ttfamily arXiv:2307.13614
  [hep-th]}}.

\bibitem{Odor:2002hk}
G.~Odor, ``{Phase transition universality classes of classical, nonequilibrium
  systems},'' \href{http://dx.doi.org/10.1103/RevModPhys.76.663}{{\em Rev. Mod.
  Phys.} {\bfseries 76} (2004) 663},
  \href{http://arxiv.org/abs/cond-mat/0205644}{{\ttfamily
  arXiv:cond-mat/0205644}}.

\bibitem{Guth:1982pn}
A.~H. Guth and E.~J. Weinberg, ``{Could the Universe Have Recovered from a Slow
  First Order Phase Transition?},''
  \href{http://dx.doi.org/10.1016/0550-3213(83)90307-3}{{\em Nucl. Phys. B}
  {\bfseries 212} (1983) 321--364}.

\bibitem{Creminelli:2008es}
P.~Creminelli, S.~Dubovsky, A.~Nicolis, L.~Senatore, and M.~Zaldarriaga, ``{The
  Phase Transition to Slow-roll Eternal Inflation},''
  \href{http://dx.doi.org/10.1088/1126-6708/2008/09/036}{{\em JHEP} {\bfseries
  09} (2008) 036}, \href{http://arxiv.org/abs/0802.1067}{{\ttfamily
  arXiv:0802.1067 [hep-th]}}.

\bibitem{branching}
T.~E. Harris, {\em The Theory of Branching Process}.
\newblock Springer-Verlag Berlin, Heidelberg, Germany, 1963.

\bibitem{bollobasriordan}
B.~Bollob\'as and O.~Riordan, ``{A simple branching process approach to the
  phase transition in $G_{n,p}$},'' {\em Electronic Journal of Combinatorics}
  {\bfseries 19} (2012) P21, \href{http://arxiv.org/abs/1207.6209}{{\ttfamily
  arXiv:1207.6209 [math.CO]}}.

\bibitem{Giudice:2021viw}
G.~F. Giudice, M.~McCullough, and T.~You, ``{Self-organised localisation},''
  \href{http://dx.doi.org/10.1007/JHEP10(2021)093}{{\em JHEP} {\bfseries 10}
  (2021) 093}, \href{http://arxiv.org/abs/2105.08617}{{\ttfamily
  arXiv:2105.08617 [hep-ph]}}.

\bibitem{Hawking:1981fz}
S.~W. Hawking and I.~G. Moss, ``{Supercooled Phase Transitions in the Very
  Early Universe},'' \href{http://dx.doi.org/10.1016/0370-2693(82)90946-7}{{\em
  Phys. Lett. B} {\bfseries 110} (1982) 35--38}.

\bibitem{Brown:1987dd}
J.~D. Brown and C.~Teitelboim, ``{Dynamical Neutralization of the Cosmological
  Constant},'' \href{http://dx.doi.org/10.1016/0370-2693(87)91190-7}{{\em Phys.
  Lett. B} {\bfseries 195} (1987) 177--182}.

\bibitem{Salem:2012wa}
M.~P. Salem, ``{Multiverse rate equation including bubble collisions},''
  \href{http://dx.doi.org/10.1103/PhysRevD.87.063501}{{\em Phys. Rev. D}
  {\bfseries 87} no.~6, (2013) 063501},
  \href{http://arxiv.org/abs/1210.7181}{{\ttfamily arXiv:1210.7181 [hep-th]}}.

\bibitem{Brown:2011ry}
A.~R. Brown and A.~Dahlen, ``{Populating the Whole Landscape},''
  \href{http://dx.doi.org/10.1103/PhysRevLett.107.171301}{{\em Phys. Rev.
  Lett.} {\bfseries 107} (2011) 171301},
  \href{http://arxiv.org/abs/1108.0119}{{\ttfamily arXiv:1108.0119 [hep-th]}}.

\bibitem{jaynes03}
E.~T. Jaynes, {\em Probability theory: The logic of science}.
\newblock Cambridge University Press, Cambridge, UK, 2003.

\bibitem{Vilenkin:1982de}
A.~Vilenkin, ``{Creation of Universes from Nothing},''
  \href{http://dx.doi.org/10.1016/0370-2693(82)90866-8}{{\em Phys. Lett. B}
  {\bfseries 117} (1982) 25--28}.

\bibitem{Hartle:1983ai}
J.~B. Hartle and S.~W. Hawking, ``{Wave Function of the Universe},''
  \href{http://dx.doi.org/10.1103/PhysRevD.28.2960}{{\em Phys. Rev. D}
  {\bfseries 28} (1983) 2960--2975}.

\bibitem{Linde:1983mx}
A.~D. Linde, ``{Quantum Creation of the Inflationary Universe},''
  \href{http://dx.doi.org/10.1007/BF02790571}{{\em Lett. Nuovo Cim.} {\bfseries
  39} (1984) 401--405}.

\bibitem{Linde:1983cm}
A.~D. Linde, ``{Quantum creation of an inflationary universe},'' {\em Sov.
  Phys. JETP} {\bfseries 60} (1984) 211--213.

\bibitem{Vilenkin:1984wp}
A.~Vilenkin, ``{Quantum Creation of Universes},''
  \href{http://dx.doi.org/10.1103/PhysRevD.30.509}{{\em Phys. Rev. D}
  {\bfseries 30} (1984) 509--511}.

\bibitem{Vilenkin:1986cy}
A.~Vilenkin, ``{Boundary Conditions in Quantum Cosmology},''
  \href{http://dx.doi.org/10.1103/PhysRevD.33.3560}{{\em Phys. Rev. D}
  {\bfseries 33} (1986) 3560}.

\bibitem{Vilenkin:1987kf}
A.~Vilenkin, ``{Quantum Cosmology and the Initial State of the Universe},''
  \href{http://dx.doi.org/10.1103/PhysRevD.37.888}{{\em Phys. Rev. D}
  {\bfseries 37} (1988) 888}.

\bibitem{Silverstein:2001xn}
E.~Silverstein, ``{(A)dS backgrounds from asymmetric orientifolds},'' {\em Clay
  Mat. Proc.} {\bfseries 1} (2002) 179,
  \href{http://arxiv.org/abs/hep-th/0106209}{{\ttfamily arXiv:hep-th/0106209}}.

\bibitem{Maloney:2002rr}
A.~Maloney, E.~Silverstein, and A.~Strominger, ``{De Sitter space in
  noncritical string theory},'' in {\em {Workshop on Conference on the Future
  of Theoretical Physics and Cosmology in Honor of Steven Hawking's 60th
  Birthday}}, pp.~570--591.
\newblock 5, 2002.
\newblock \href{http://arxiv.org/abs/hep-th/0205316}{{\ttfamily
  arXiv:hep-th/0205316}}.

\bibitem{DeLuca:2021pej}
G.~B. De~Luca, E.~Silverstein, and G.~Torroba, ``{Hyperbolic compactification
  of M-theory and de Sitter quantum gravity},''
  \href{http://dx.doi.org/10.21468/SciPostPhys.12.3.083}{{\em SciPost Phys.}
  {\bfseries 12} no.~3, (2022) 083},
  \href{http://arxiv.org/abs/2104.13380}{{\ttfamily arXiv:2104.13380
  [hep-th]}}.

\bibitem{Katz1953}
L.~Katz, ``A new status index derived from sociometric analysis,'' {\em
  Psychometrika} {\bfseries 18} (1953) 39--43.

\bibitem{Lee:1987qc}
K.-M. Lee and E.~J. Weinberg, ``{Decay of the True Vacuum in Curved
  Space-time},'' \href{http://dx.doi.org/10.1103/PhysRevD.36.1088}{{\em Phys.
  Rev. D} {\bfseries 36} (1987) 1088}.

\bibitem{Dyson:2002pf}
L.~Dyson, M.~Kleban, and L.~Susskind, ``{Disturbing implications of a
  cosmological constant},''
  \href{http://dx.doi.org/10.1088/1126-6708/2002/10/011}{{\em JHEP} {\bfseries
  10} (2002) 011}, \href{http://arxiv.org/abs/hep-th/0208013}{{\ttfamily
  arXiv:hep-th/0208013}}.

\bibitem{Farhi:1989yr}
E.~Farhi, A.~H. Guth, and J.~Guven, ``{Is It Possible to Create a Universe in
  the Laboratory by Quantum Tunneling?},''
  \href{http://dx.doi.org/10.1016/0550-3213(90)90357-J}{{\em Nucl. Phys. B}
  {\bfseries 339} (1990) 417--490}.

\bibitem{Fischler:1989se}
W.~Fischler, D.~Morgan, and J.~Polchinski, ``{Quantum Nucleation of False
  Vacuum Bubbles},'' \href{http://dx.doi.org/10.1103/PhysRevD.41.2638}{{\em
  Phys. Rev. D} {\bfseries 41} (1990) 2638}.

\bibitem{Fischler:1990pk}
W.~Fischler, D.~Morgan, and J.~Polchinski, ``{Quantization of False Vacuum
  Bubbles: A Hamiltonian Treatment of Gravitational Tunneling},''
  \href{http://dx.doi.org/10.1103/PhysRevD.42.4042}{{\em Phys. Rev. D}
  {\bfseries 42} (1990) 4042--4055}.

\bibitem{DeAlwis:2019rxg}
S.~P. De~Alwis, F.~Muia, V.~Pasquarella, and F.~Quevedo, ``{Quantum Transitions
  Between Minkowski and de Sitter Spacetimes},''
  \href{http://dx.doi.org/10.1002/prop.202000069}{{\em Fortsch. Phys.}
  {\bfseries 68} no.~9, (2020) 2000069},
  \href{http://arxiv.org/abs/1909.01975}{{\ttfamily arXiv:1909.01975
  [hep-th]}}.

\bibitem{Fu:2019oyc}
Z.~Fu and D.~Marolf, ``{Bag-of-gold spacetimes, Euclidean wormholes, and
  inflation from domain walls in AdS/CFT},''
  \href{http://dx.doi.org/10.1007/JHEP11(2019)040}{{\em JHEP} {\bfseries 11}
  (2019) 040}, \href{http://arxiv.org/abs/1909.02505}{{\ttfamily
  arXiv:1909.02505 [hep-th]}}.

\bibitem{Mirbabayi:2020grb}
M.~Mirbabayi, ``{Uptunneling to de Sitter},''
  \href{http://dx.doi.org/10.1007/JHEP09(2020)070}{{\em JHEP} {\bfseries 09}
  (2020) 070}, \href{http://arxiv.org/abs/2003.05460}{{\ttfamily
  arXiv:2003.05460 [hep-th]}}.

\bibitem{Olum:2021pux}
K.~D. Olum, P.~Upadhyay, and A.~Vilenkin, ``{Black holes and uptunneling
  suppress Boltzmann brains},''
  \href{http://dx.doi.org/10.1103/PhysRevD.104.023528}{{\em Phys. Rev. D}
  {\bfseries 104} no.~2, (2021) 023528},
  \href{http://arxiv.org/abs/2105.00457}{{\ttfamily arXiv:2105.00457
  [hep-th]}}.

\bibitem{DAGpaper}
B.~Karrer and M.~E.~J. Newman, ``Random graph models for directed acyclic
  networks,'' {\em Physical Review E} {\bfseries 80} (2009) 046110,
  \href{http://arxiv.org/abs/0907.4346}{{\ttfamily arXiv:0907.4346
  [physics.soc-ph]}}.

\bibitem{funnelfig}
A.~Samarakoon, T.~J. Sato, T.~Chen, G.-W. Chern, J.~Yang, I.~Klich,
  R.~Sinclair, H.~Zhou, and S.-H. Lee, ``{Aging, memory, and nonhierarchical
  energy landscape of spin jam},'' {\em Proc. Natl. Acad. Sci.} {\bfseries 113}
  (2016) 11806--11810, \href{http://arxiv.org/abs/1707.03086}{{\ttfamily
  arXiv:1707.03086 [cond-mat.dis-nn]}}.

\bibitem{percolationtextbook}
D.~Stauffer and A.~Aharony, {\em Introduction to Percolation Theory}.
\newblock Taylor \& Francis, London, UK, 2003.

\bibitem{Essamreview}
J.~W. Essam, ``{Percolation theory},'' {\em Rep. Prog. Phys.} {\bfseries 43}
  (1980) 833.

\bibitem{bowtie}
A.~Broder, R.~Kumar, F.~Maghoul, P.~Raghavan, S.~Rajagopalan, R.~Stata,
  A.~Tomkins, and J.~Wiener, ``{Graph structure in the web},'' {\em Computer
  Networks} {\bfseries 33} (2000) 309.

\bibitem{SF1}
A.-L. Barab\'asi and R.~Albert, ``{Emergence of scaling in random networks},''
  {\em Science.} {\bfseries 286} (1999) 509,
  \href{http://arxiv.org/abs/cond-mat/9910332}{{\ttfamily
  arXiv:cond-mat/9910332}}.

\bibitem{SFreview}
R.~Albert and A.-L. Barab\'asi, ``{Statistical mechanics of complex
  networks},'' {\em Rev. Mod. Phys.} {\bfseries 74} (2002) 47,
  \href{http://arxiv.org/abs/cond-mat/0106096}{{\ttfamily
  arXiv:cond-mat/0106096}}.

\bibitem{scalefreeexponents}
R.~Cohen, D.~ben Avraham, and S.~Havlin, ``{Percolation Critical Exponents in
  Scale-Free Networks},'' {\em Phys. Rev. E} {\bfseries 66} (2002) 036113,
  \href{http://arxiv.org/abs/cond-mat/0202259}{{\ttfamily
  arXiv:cond-mat/0202259}}.

\bibitem{scalefreepercolation}
N.~Schwartz, R.~Cohen, D.~ben Avraham, A.-L. Barab\'asi, and S.~Havlin,
  ``{Percolation in directed scale-free networks},'' {\em Phys. Rev. E}
  {\bfseries 66} (2002) 015104,
  \href{http://arxiv.org/abs/cond-mat/0204523}{{\ttfamily
  arXiv:cond-mat/0204523}}.

\bibitem{BogunaSerrano}
M.~Boguna and M.~A. Serrano, ``{Generalized percolation in random directed
  networks},'' {\em Phys. Rev. E} {\bfseries 72} (2005) 016106,
  \href{http://arxiv.org/abs/cond-mat/0501533}{{\ttfamily
  arXiv:cond-mat/0501533}}.

\bibitem{randomweightedgraph}
D.~Garlaschelli, ``{The weighted random graph model},'' {\em New Journal of
  Physics} {\bfseries 11} (2009) 073005,
  \href{http://arxiv.org/abs/0902.0897}{{\ttfamily arXiv:0902.0897
  [cond.mat]}}.

\bibitem{Sumitomo:2012wa}
Y.~Sumitomo and S.-H. Tye, ``{A Stringy Mechanism for A Small Cosmological
  Constant},'' \href{http://dx.doi.org/10.1088/1475-7516/2012/08/032}{{\em
  JCAP} {\bfseries 08} (2012) 032},
  \href{http://arxiv.org/abs/1204.5177}{{\ttfamily arXiv:1204.5177 [hep-th]}}.

\bibitem{Sumitomo:2012vx}
Y.~Sumitomo and S.~H. Tye, ``{A Stringy Mechanism for A Small Cosmological
  Constant - Multi-Moduli Cases -},''
  \href{http://dx.doi.org/10.1088/1475-7516/2013/02/006}{{\em JCAP} {\bfseries
  02} (2013) 006}, \href{http://arxiv.org/abs/1209.5086}{{\ttfamily
  arXiv:1209.5086 [hep-th]}}.

\bibitem{Sumitomo:2012cf}
Y.~Sumitomo and S.-H. Tye, ``{Preference for a Vanishingly Small Cosmological
  Constant in Supersymmetric Vacua in a Type IIB String Theory Model},''
  \href{http://dx.doi.org/10.1016/j.physletb.2013.05.027}{{\em Phys. Lett. B}
  {\bfseries 723} (2013) 406--410},
  \href{http://arxiv.org/abs/1211.6858}{{\ttfamily arXiv:1211.6858 [hep-th]}}.

\bibitem{Danielsson:2012by}
U.~Danielsson and G.~Dibitetto, ``{On the distribution of stable de Sitter
  vacua},'' \href{http://dx.doi.org/10.1007/JHEP03(2013)018}{{\em JHEP}
  {\bfseries 03} (2013) 018}, \href{http://arxiv.org/abs/1212.4984}{{\ttfamily
  arXiv:1212.4984 [hep-th]}}.

\bibitem{Sumitomo:2013vla}
Y.~Sumitomo, S.~Tye, and S.~S. Wong, ``{Statistical Distribution of the Vacuum
  Energy Density in Racetrack K{\"a}hler Uplift Models in String Theory},''
  \href{http://dx.doi.org/10.1007/JHEP07(2013)052}{{\em JHEP} {\bfseries 07}
  (2013) 052}, \href{http://arxiv.org/abs/1305.0753}{{\ttfamily arXiv:1305.0753
  [hep-th]}}.

\bibitem{Tye:2016jzi}
S.~H.~H. Tye and S.~S.~C. Wong, ``{Linking Light Scalar Modes with A Small
  Positive Cosmological Constant in String Theory},''
  \href{http://dx.doi.org/10.1007/JHEP06(2017)094}{{\em JHEP} {\bfseries 06}
  (2017) 094}, \href{http://arxiv.org/abs/1611.05786}{{\ttfamily
  arXiv:1611.05786 [hep-th]}}.

\bibitem{Rao_Caflisch_2004}
F.~Rao and A.~Caflisch, ``{The protein folding network},'' {\em J. Mol. Biol.}
  {\bfseries 342} (2009) 299--306,
  \href{http://arxiv.org/abs/q-bio/0403034}{{\ttfamily arXiv:q-bio/0403034
  [q-bio.BM]}}.

\bibitem{wilf}
H.~S. Wilf, {\em generatingfunctionology}.
\newblock AK Peters/CRC; 3rd edition, Natick, MA, USA, 2006.

\end{thebibliography}\endgroup
\end{document}